\documentclass[Afour,sageh,times]{sagej}

\usepackage[T1]{fontenc}
\usepackage[utf8x]{inputenc}
\usepackage{lmodern}
\usepackage[polutonikogreek,english]{babel}
\usepackage[autostyle]{csquotes}
\usepackage{ifdraft}
\usepackage{textgreek}
\usepackage{graphicx}

\usepackage[colorlinks,bookmarksopen,bookmarksnumbered,citecolor=red,urlcolor=red]{hyperref}
\usepackage{moreverb}
\usepackage[disable]{todonotes}
\usepackage{url}
\usepackage{cleveref}
\usepackage[perpage]{footmisc}
\usepackage{xcolor}
\usepackage{listings}
\usepackage[charsperline=55]{jlcode}

\newcommand\BibTeX{{\rmfamily B\kern-.05em \textsc{i\kern-.025em b}\kern-.08em
T\kern-.1667em\lower.7ex\hbox{E}\kern-.125emX}}

\def\volumeyear{2022}

\begin{document}

\runninghead{Churavy \textit{et~al.}}


\title{Bridging HPC Communities through the Julia Programming Language}

\author{%
Valentin Churavy\affilnum{1},
William F Godoy\affilnum{2},
Carsten Bauer\affilnum{3},
Hendrik Ranocha\affilnum{4},
Michael Schlottke-Lakemper\affilnum{5,6},
Ludovic R\"ass\affilnum{7,8},
Johannes Blaschke\affilnum{9},
Mosè Giordano\affilnum{10},
Erik Schnetter\affilnum{11,12,13},
Samuel Omlin\affilnum{14},
Jeffrey S. Vetter\affilnum{2},
Alan Edelman\affilnum{1}
}

\affiliation{\\[-3ex]
\affilnum{1} Massachussetts Institute of Technology, USA\\
\affilnum{2} Oak Ridge National Laboratory, USA\\
\affilnum{3} Paderborn Center for Parallel Computing, Paderborn University, Germany\\
\affilnum{4} Department of Mathematics, University of Hamburg, Germany\\
\affilnum{5} Applied and Computational Mathematics, RWTH Aachen University, Germany\\
\affilnum{6} High-Performance Computing Center Stuttgart (HLRS), University of Stuttgart, Germany\\
\affilnum{7} Laboratory of Hydraulics, Hydrology and Glaciology (VAW), ETH Zurich, Switzerland\\
\affilnum{8} Swiss Federal Institute for Forest, Snow and Landscape Research (WSL), Birmensdorf, Switzerland\\
\affilnum{9} National Energy Research Scientific Computing Center, Lawrence Berkeley National Laboratory, 1 Cyclotron Road, Berkeley, CA 94720, USA \\
\affilnum{10} Centre for Advanced Research Computing, University College London, Gower Street, London, WC1E 6BT, United Kingdom\\
\affilnum{11} Perimeter Institute, 31 Caroline St. N., Waterloo, ON, Canada N2L 2Y5\\
\affilnum{12} Department of Physics and Astronomy, University of Waterloo, 200 University Avenue West, Waterloo, Ontario, Canada N2L 3G1\\
\affilnum{13} Center for Computation \& Technology, Louisiana State University, Baton Rouge, LA 70803, USA\\
\affilnum{14} Swiss National Supercomputing Centre (CSCS), ETH Zurich, Switzerland
}

\corrauth{Valentin Churavy,
Computer Science and Artificial Intelligence Laboratory,
Massachusetts Institute of Technology,
Cambridge, Massachusetts,
USA}

\email{vchuravy@mit.edu}

\date{2022-11-01}

\begin{abstract}
The Julia programming language has evolved into a modern alternative to fill existing gaps in scientific computing and data science applications.
Julia leverages a unified and coordinated single-language and ecosystem paradigm and has a proven track record of achieving high performance without sacrificing user productivity. These aspects make Julia a viable  alternative to high-performance computing's (HPC's) existing and increasingly costly many-body workflow composition strategy in which traditional HPC languages (e.g., Fortran, C, C\texttt{++}) are used for simulations, and higher-level languages (e.g., Python, R, MATLAB) are used for data analysis and interactive computing. Julia's rapid growth in language capabilities, package ecosystem, and community make it a promising universal language for HPC.
This paper presents the views of a multidisciplinary group of researchers from academia, government, and industry that advocate for an HPC software development paradigm that emphasizes developer productivity, workflow portability, and low barriers for entry. We believe that the Julia programming language, its ecosystem, and its community provide modern and powerful capabilities that enable this group's objectives. Crucially, we believe that Julia can provide a feasible and less costly approach to programming scientific applications and workflows that target HPC facilities.
In this work, we examine the current practice and role of Julia as a common, end-to-end programming model to address major challenges in scientific reproducibility, data-driven AI/machine learning, co-design and workflows, scalability and performance portability in heterogeneous computing, network communication, data management, and community education. As a result, the diversification of current investments to fulfill the needs of the upcoming decade is crucial as more supercomputing centers prepare for the exascale era.
\end{abstract}

\keywords{High-Performance Computing, HPC, Julia, Programming Language, Workflows, Productivity, Performance Portability}

\maketitle


\section{Introduction}

The Julia programming language~\citep{bezanson2018julia} was designed in the last decade to be a novel, high-level, dynamic, and high-performance approach to numerical computing. Julia programs compile as efficient native code for several heterogeneous architectures via the open-source LLVM compiler~\citep{1281665}. The syntax builds upon the success of Fortran for multidimensional arrays and mathematical abstractions~\citep{4038201} and combines with a rich ecosystem that includes high-level interfaces for data structures, analysis, visualization, AI frameworks, and interactive computing.
Julia was also designed to address aspects that are typically offloaded to a language ecosystem but are still necessary in the overall scientific discovery process 
(e.g., reproducibility, packaging, environment portability). 
Julia also includes a powerful macros system for code instrumentation, interactive computing capabilities, and lightweight interoperability with existing C and Fortran codes---especially highly optimized high-performance computing (HPC) software frameworks and libraries. 
Julia offers a powerful workflow composition strategy because existing highly optimized HPC frameworks can be combined seamlessly with high-performance Julia kernel code for computation and data management on heterogeneous systems. This creates a powerful synergy for programming HPC systems as more emphasis is placed on performance portability and programmer productivity in the overall workflow process, beyond simulations~\citep{9309042}.

Software development that targets HPC facilities for scientific discovery is a nontrivial and highly specialized task~\citep{10.1007/978-3-0348-8534-8_11}. 
Efficient use of HPC facilities for computational science and engineering (CSE) is a multidisciplinary orchestration among several stakeholders. This process requires intimate knowledge of the application's target domain, the targeted system's architecture, and the algorithms in the frameworks and libraries that handle the scalable computation, communication, and data performance aspects within the co-design process. 
As we reach the physical limits of Moore's Law in semiconductor technology~\citep{moore1998cramming,7368023}, several heterogeneous architectures and programming models have emerged~\citep{osti_1473756} during a time in which the first exascale systems are being deployed for the HPC community.
On the software technology side, major vendors have converged around the LLVM open-source project~\citep{1281665} as the back-end technology of choice for their plethora of compilers and programming models. LLVM's modularity, reusability, and platform-agnostic intermediate representation (IR) enables the desired productivity and performance portability characteristics. 
At the same time, custom hardware accelerators are powering the computational demands associated with AI applications at a wide range of scales. 
Consequently, the current landscape offers unique opportunities to rethink traditional HPC aspects such as end-to-end co-design for performance portability of complex workflows, large-scale rapid prototyping, and collaboration with dominant cloud and mobile computing ecosystems~\citep{https://doi.org/10.48550/arxiv.2203.02544}.

The present work outlines our view that Julia can challenge the current status quo---in which high-level languages designed with productivity in mind cannot easily achieve the desired levels of performance---while also reducing the costs associated with the learning curve, implementation, and maintenance of an infrastructure based on compiled HPC languages.
Much of Fortran's success can be attributed to providing an answer to the original question ~\citep{backus1980programming}: ``Can a machine translate a sufficiently rich mathematical language into a sufficiently economical program at a sufficiently low cost to make the whole affair feasible?'' Julia attempts to solve a similar technical and economical challenge according to the current landscape by expanding on the traditional HPC focus of simulation performance towards workflow applications. 
Just like Fortran has been the dominant language for science in the last several decades, Julia can be seen as a unifying domain-specific language (DSL) for science that targets modern HPC requirements for simulations, data analysis, workflows, and interactive computing.
The expected return on investment for leveraging Julia is an increase in productivity when addressing the end-to-end co-design needs of multidisciplinary HPC projects, without a drop performance portability, while also keeping development in a single unifying language and ecosystem. 
The latter is particularly important in the convergence of AI~+~HPC workflows for science as AI has been one of the primary drivers in computational sciences in the past decade~\citep{osti_1604756}. 

The rest of the paper describes what makes the Julia language an attractive investment for scientific discovery with HPC. Section~\ref{sec:background} provides background information on the history and efforts around programming languages for HPC, including initiatives that led to the proliferation of current programming models. 
Section~\ref{sec:community} describes the community adoption, interest in leadership facilities around the world, and the package development and deployment process to enable reproducible science at those centers. Section~\ref{sec:teaching} outlines the value of Julia as a first language for teaching HPC concepts. Performance and scalability, which are key aspects of HPC's ethos, are described in Section~\ref{sec:scalability-portability}, including experiences in heterogeneous architectures that combine the power of CPUs and GPUs (graphics processing units). Section~\ref{sec:julia success stories} presents an overview of Julia success stories, including recent research studies that describe performance aspects and community adoption in the broader field of CSE.
Section~\ref{sec:interoperability} describes the central aspect of Julia's interoperability with C and Fortran that allows access to highly optimized HPC frameworks, along with reusability with Python's existing frameworks, for a powerful workflow composability strategy. Section~\ref{sec:conclusions} summarizes our conclusions and vision for Julia and potential opportunities and investments for the HPC community.

\section{Background}
\label{sec:background}

The development of programming languages for HPC has a rich and varied history. Early on, the needs of HPC and mainstream computing were mostly aligned around number crunching for numerical calculations, which led to the development of Fortran~\citep{4038201} as the first high-level HPC language in the 1950s. To this day, Fortran continues strongly as a leading programming language for HPC owing to its legacy of investments and highly optimized implementations~\citep{9736688}. As computing evolved and added more requirements at the system level to perform data movement, parallel processing, analysis, and visualization, C~\citep{kernighan1988c} and C\texttt{++}~\citep{stroustrup2013c++} became the dominant system-level and numerical computing languages in HPC.

At the beginning of the 21st century, the Defense Advanced Research Projects Agency's (DARPA's) High-Productivity Computing Systems (HPCS) program~\citep{dongarra2008darpa} described the common practice for HPC software as writing kernels in a compiled sequential language (e.g., Fortran, C, C\texttt{++}) and then parallelizing them in a memory-distributed model based on the standard Message Passing Interface (MPI)~\citep{gropp1999using}. HPCS funded an effort to develop new programming languages that targeted productivity, and this resulted in Cray's Chapel Parallel Programming Language~\citep{doi:10.1177/1094342007078442}, IBM's X10~\citep{10.1145/1229428.1229483}, and Sun's Fortress~\citep{allen2005fortress}. Other efforts included those based on Fortran and C extensions, such as Coarray Fortran~\citep{10.1145/289918.289920} and the unified parallel C~\citep{el2005upc}. In general, these new programming languages offered an alternative to traditional message passing and multithreaded programming models by using approaches such as partitioned global address space~\citep{el2005upc,Almasi2011}.

The past decade has seen several disruptive trends that led to the current landscape of extreme heterogeneity: (1) the emergence and adoption of GPU computing as a disruptive technology in HPC~\citep{5289128} owing to its performance, programmability, and energy efficiency~\citep{5598297}; (2) the flattening of Moore's Law in the CMOS technology manufacturing industry; and (3) the adoption of LLVM as the compiler of choice from major vendors. These trends have led to the proliferation of new standardized, vendor-specific, and third-party programming models in the past decade. These models target HPC languages used to manage the increased heterogeneity of contemporary systems: OpenCL~\citep{munshi2009opencl}, CUDA~\citep{buck2007gpu}, HIP~\citep{rocm_hip}, OpenMP~\citep{openmp}, OpenACC~\citep{wienke2012openacc}, SYCL~\citep{reyes2016sycl}, Kokkos~\citep{Kokkos}, and RAJA~\citep{raja} among others.

Overall, programming languages used in HPC are not specifically designed for science, with Fortran being the exception. 
This has been a sustainable model owing to vendor and community support, especially for C\texttt{++} and Python as rapidly evolving general-purpose languages. The HPC software stacks funded by the US Department of Energy's (DOE's) Exascale Computing Project (ECP)~\citep{osti_1463232,heroux2019extreme,doi:10.1177/1094342010391989} have continued to build upon the legacy of Fortran, C, and C\texttt{++}, and Python's high-productivity ecosystem has been widely adopted for data analysis, AI, and workflow composition~\citep{9307940}.
\citet{Ousterhout1998-ac} already observed the split of programming languages into two distinct groups: \emph{implementation} and \emph{scripting}. It was anticipated that scripting language interfaces that glue together the underlying system components would become a dominant model with trade-offs and challenges of its own. A major challenge is the bifurcation of the different communities and the high cost for learning and maintaining multiple technologies and ecosystems.
This is even more noticeable in the era of AI because frameworks such as TensorFlow~\citep{tensorflow}, PyTorch~\citep{paszke2019pytorch}, JAX~\citep{jax}, and Firedrake~\citep{Firedrake} target end users in high-productivity languages.
Closing the gaps between HPC's needs and ease of use is a nontrivial effort that adds overheads costs~\citep{9678726,7836841}. 

Julia was designed to prioritize research and development cycles from idea to performance portability for scientific discovery. Reducing the overhead development costs in this landscape is crucial as future systems become more complex and heterogeneous. The unified language approach builds upon the requirements of the scientific communities that are facing these challenges. In this regard, Julia has attracted domain scientists and practitioners from multiple disciplines to create a community that continues to grow and establish synergistic collaborations. We propose that Julia is a sustainable investment for HPC software projects as future challenges continue to add costs to the scientific discovery objectives that drive and justify the large strategic investments in these systems.

\section{Community}
\label{sec:community}
The Julia language community is made up of many people working in various scientific and technical domains, and even the original Julia manifesto\footnote{\url{https://julialang.org/blog/2012/02/why-we-created-julia/}, accessed 08-16-2022.} described the target demographic as including scientific computing, machine learning, data mining, large-scale linear algebra, and distributed and parallel computing. The umbrella term for these domains is \textit{technical computing}.

The original developers of Julia aimed to design an open-source language to tackle problems in technical computing, and from there the community has grown
to encompass a wide variety of use cases---from web servers, to databases, to numerical simulations on HPC systems. Although Julia is now recognized as a general-purpose programming language, the early focus on technical computing is still apparent. Common challenges for people working in technical computing are reproducibility and software distribution, and we will discuss these problems in \Cref{subsec:reproducibility}. The rest of this section focuses on the HPC subdemographic of the Julia community (\Cref{subsec:community-hpc}), Julia at the National Energy Research Scientific Computing Center~(NERSC) (\Cref{subsec:community-nersc}), and the HPC centers around the world (\Cref{subsec:community-hpccenters}).

\subsection{Package development and reproducibility}
\label{subsec:reproducibility}

Julia was specifically designed to fulfill the Fortran dream of automating the translation of formulas into efficient executable code~\citep{Bezanson2017-ca}.  Additionally, Julia addresses the two-language problem by closing the gap between developers and users of scientific software.  This is achieved with an intuitive language and by providing users with tools to more easily follow good, modern programming practices---including documentation, testing, and continuous integration.  A recent survey of the packages collected in the General registry showed a strong adoption of these practices: over 95\% of packages had tests and ran them with continuous integration services, and almost 90\% of packages had documentation~\citep{hanson:general}.  The adoption of these practices is also made simpler by package templates such as those provided by \texttt{PkgTemplates.jl}~\citep{pkgtemplates_jl}.

Building on the experience of other languages, Julia comes with a built-in package manager, \texttt{Pkg.jl}, which can install packages and manage package environments similar to the concept of virtual environments in Python.  Julia package environments are defined by two text files: \texttt{Project.toml} and \texttt{Manifest.toml}. \texttt{Project.toml} specifies the list of direct dependencies of an environment and their compatibility constraints. \texttt{Manifest.toml} captures all direct and indirect dependencies of the environment and uses the appropriate versions of each software module for the present environment.  When both files are provided, they fully define a computational environment, and this environment can then be recreated later or on a different machine.
We use these features in the reproducibility repository described in this
paper~\citep{churavy2022bridgingRepro}.

Julia packages are set up as Git repositories that can be hosted on any Git hosting services.  Many development tools, including continuous integration tools and online package documentation solutions, are well integrated with GitHub and GitLab, which are the two most popular repository hosting services within the Julia community.  All versions of packages recorded in the General registry are automatically duplicated by the servers used by \texttt{Pkg.jl} to prevent deleted packages from taking their dependents out with them---an unfortunate scenario that played out with the \texttt{left-pad} JavaScript package~\citep{theregister:left-pad}.

Julia allows for writing an entire software stack in a single language thanks to its unique combination of ease-of-use and speed. However, Julia users often want to use legacy code already written in other languages, such as C, C\texttt{++}, Fortran, Python, or R.  Julia offers the capability to call functions in shared libraries written in C and Fortran and libraries written in any other languages that provide a C-like interface. Third-party packages such as \texttt{Clang.jl}~\citep{clang_jl} and \texttt{CBinding.jl}~\citep{cbinding_jl} enable the automatic creation of Julia bindings for C libraries by parsing their header files.  Some packages enable other languages to be used directly from within a Julia process, including but not limited to \texttt{PyCall.jl}~\citep{pycall_jl} and \texttt{PythonCall.jl}~\citep{PythonCall.jl} for Python, \texttt{RCall.jl}~\citep{rcall_jl} for R, and \texttt{MATLAB.jl}~\citep{matlab_jl} for MATLAB. \texttt{CxxWrap.jl}~\citep{cxxwrap_jl} makes it possible to interface C\texttt{++} shared libraries by using a static binding generator.

Within the Julia ecosystem, binary libraries and executables are usually managed with \texttt{BinaryBuilder.jl}~\citep{binaybuilder_jl}. This framework allows package developers to compile pre-built versions of the binaries for all Julia-supported platforms and then upload them to GitHub. The corresponding and automatically generated packages, called \textit{JLLs}, provide a programmatic interface to call into libraries or run executables. The JLLs are regular Julia packages that, when installed, automatically download the corresponding libraries or executables, thus relieving users from the effort of installing or compiling external libraries themselves. That the JLLs are regular Julia packages also means that they can be recorded in the package environment, thus extending the reproducibility of a computing environment to libraries and programs in other languages. The \texttt{BinaryBuilder.jl} framework is usually seen as successful because it provides straightforward handling of external libraries in the general cases. This may cause some friction in HPC settings in which users would like to leverage the system's fine-tuned libraries. However there are mechanisms to override the pre-built libraries provided by JLL packages while still using their programmatic interface.

\subsection{Uptake of Julia in the HPC community}
\label{subsec:community-hpc}

As Julia places performance at the core of the language, the HPC community has been among the early adopters of the Julia language. Notable examples of early HPC readiness are the petaflop runs at DOE's NERSC~\citep{hpcwire:celeste}.
The Celeste Julia code, which analyzes astronomical images, achieved 1.54 petaflops using 1.3 million threads on 9,300 Knights Landing (KNL) nodes of the Cori supercomputer.
At the time, this represented an important milestone because experimental and observational science workflows are typically coded using high-productivity interpreted languages that are optimized for rapid prototyping but not for performance.
These scientific domains have some of the highest adoption rates for Julia and rely on rapid prototyping, complex workflows, and interactive computing.

\subsection{A detailed look at Julia use at NERSC}
\label{subsec:community-nersc}

NERSC is a DOE user facility with approximately 8,000 users. Most users are employed at universities and DOE laboratories, and half are early career scientists, including graduate students and postdocs. Projects using NERSC's HPC systems are funded by DOE program offices: Basic Energy Sciences, High-Energy Physics, Biological and Environmental Research, Fusion Energy Sciences, Nuclear Physics, Advanced Computing Research, and Small Business Innovation Research. Owing to this breadth of research, a survey of NERSC users provides insights into a broad research community.

\begin{figure}
    \centering
    \includegraphics[width=0.35\textwidth]{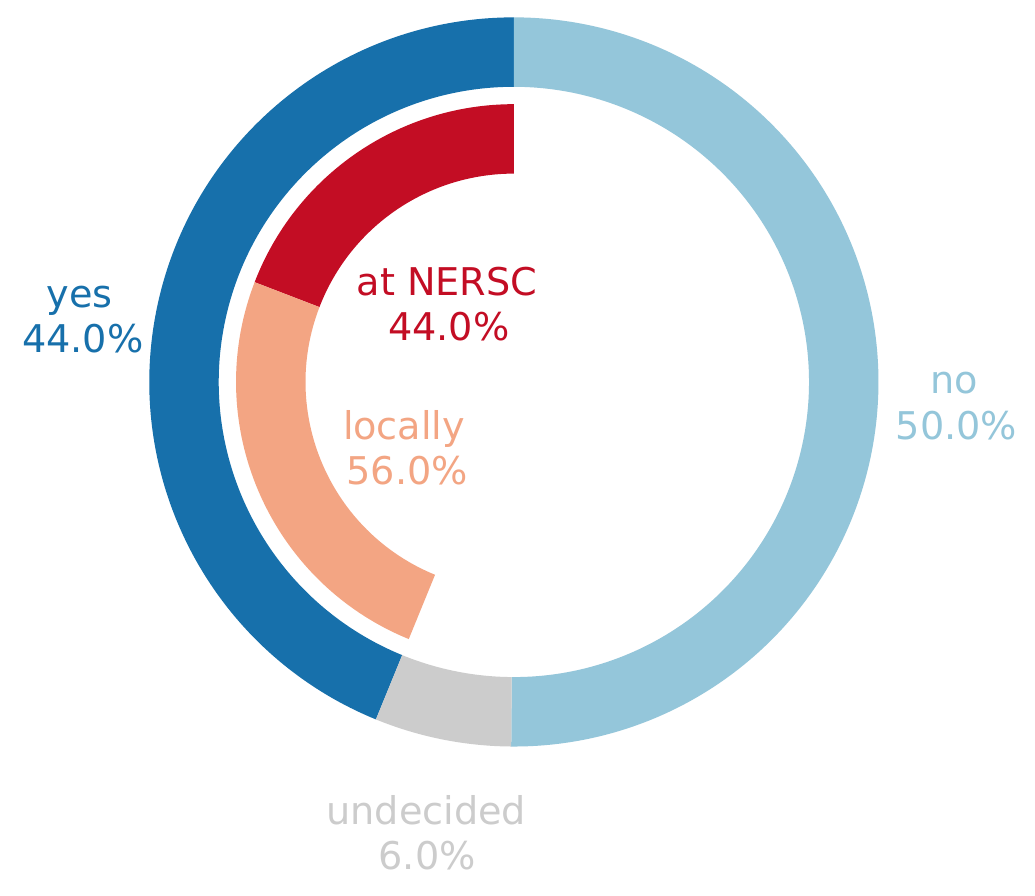}
    \caption{NERSC user survey: 44\% of all respondents (415 NERSC users) plan to use Julia in the future. Of those, 44\% plan to use Julia at NERSC.}
    \label{fig:planned_use}
\end{figure}

NERSC monitors the use of the {\tt module load julia} command (among many others) with MODS (Monitoring of Data Systems). MODS captures workflows that use NERSC's official Julia install---users that install their own version of Julia are not tracked. MODS reports that 132 unique, non-staff users loaded a Julia module at least once in 2021. MODS also shows a gradual increase in Julia module usage at NERSC, but this view is limited. To see a clearer picture of the community's future plans, we surveyed NERSC users and received 415 responses. Most responded within the first 2 days, thereby indicating strong interest. The survey results showed that 44\% of respondents are planning to use Julia (\Cref{fig:planned_use}).

\subsection{User support and interest at major HPC centers}
\label{subsec:community-hpccenters}

Julia is supported by several major HPC centers surveyed in the United States and Europe (see~\Cref{tab:center-support}). Official support at HPC centers takes the form of (1) inclusion of Julia and possibly packages in the official module tree; (2) site-specific configurations (e.g., MPI, I/O); (3) official user documentation; and (4) support for user trouble tickets.

\begin{table*}
    \center
    \begin{tabular}{p{0.20\linewidth}p{0.16\linewidth}p{0.01\linewidth}p{0.01\linewidth}p{0.01\linewidth}p{0.01\linewidth}p{0.21\linewidth}p{0.21\linewidth}}
    \toprule
        \textbf{Center Name} &
        \textbf{System Names} &
        \multicolumn{4}{c}{\hspace{-1Em}\begin{tabular}[t]{l}\textbf{Support Level}\end{tabular}} &
        \textbf{CPU Architecture} &
        \textbf{Accelerators}
        \\
        && P & U & I & D &&
        \\

        \midrule
        Australasia & & & & & & & \\
        \midrule

        NeSI &
        \hspace{-0.6em}\begin{tabular}[lt]{l}Mahuika,\\Māui\end{tabular} &
        \checkmark & \checkmark & \checkmark & \checkmark &
        \hspace{-0.6em}\begin{tabular}[t]{l}Intel Broadwell,\\Intel Cascade Lake,\\AMD Milan\end{tabular} &
        \hspace{-0.6em}\begin{tabular}[t]{l}NVIDIA P100,\\NVIDIA P100\end{tabular}
        \\
        
        \midrule
        Europe & & & & & & & \\
        
        \midrule
        
        ARC (UCL) &
        \hspace{-0.6em}\begin{tabular}[t]{l}Myriad,\\Kathleen,\\Michael,\\Young\end{tabular} &
        \checkmark & \checkmark &  & \checkmark &
        Various Intel Xeon &
        Various GPUs
        \\

        CSC (EuroHPC) &
        LUMI &
        \checkmark & \checkmark &  & \checkmark &
        AMD Milan &
        AMD M250X
        \\

        CSCS &
        Piz Daint &
        \checkmark & \checkmark & \checkmark & \checkmark &
        \hspace{-0.6em}\begin{tabular}[t]{l}Intel Broadwell,\\Intel Haswell\end{tabular} &
        NVIDIA P100
        \\

        DESY IT &
        Maxwell &
        \checkmark &  & \checkmark & \checkmark &
        \hspace{-0.6em}\begin{tabular}[t]{l}Various AMD Epyc\\Various Intel Xeon\end{tabular} &
        Various GPUs
        \\

        HLRS &
        Hawk &
        \checkmark & \checkmark & \checkmark & \checkmark &
        AMD Rome &
        NVIDIA A100
        \\

        HPC2N (Umeå) &
        Kebnekaise &
        \checkmark & \checkmark &  & \checkmark & 
        \hspace{-0.6em}\begin{tabular}[t]{l}Intel Broadwell,\\Intel Skylake\end{tabular} &
        \hspace{-0.6em}\begin{tabular}[t]{l}NVIDIA K80,\\NVIDIA V100\end{tabular}
        \\

        IT4I (EuroHPC) &
        Karolina &
        \checkmark & \checkmark & \checkmark & \checkmark &
        AMD Rome &
        NVIDIA A100
        \\

        IZUM (EuroHPC) &
        Vega &
        \checkmark & \checkmark & \checkmark & \checkmark &
        AMD Rome &
        NVIDIA A100
        \\

        LuxProvide (EuroHPC) &
        MeluXina &
        \checkmark &  & \checkmark & \checkmark &
        AMD Rome &
        NVIDIA A100
        \\

        PC2 (Paderborn) &
        Noctua 1 &
        \checkmark & \checkmark & \checkmark & \checkmark &
        Intel Skylake &
        \hspace{-0.6em}\begin{tabular}[t]{l}Various GPUs\end{tabular}
        \\

        PC2 (Paderborn) &
        Noctua 2 &
        \checkmark & \checkmark & \checkmark & \checkmark &
        AMD Milan &
        \hspace{-0.6em}\begin{tabular}[t]{l}NVIDIA A100,\\Xilinx U280,\\Intel Stratix 10\end{tabular}
        \\

        \hspace{-0.6em}\begin{tabular}[t]{l}ULHPC\\(Luxembourg)\end{tabular} &
        \hspace{-0.6em}\begin{tabular}[t]{l}Aion,\\Iris\end{tabular} &
        \checkmark &  & \checkmark & \checkmark &
        \hspace{-0.7em}\begin{tabular}[t]{l}AMD Rome,\\Intel Broadwell,\\Intel Skylake\end{tabular} &
        NVIDIA V100
        \\

        ZDV (Mainz) &
        MOGON II &
        \checkmark &  &  & \checkmark &
        \hspace{-0.6em}\begin{tabular}[t]{l}Intel Broadwell,\\Intel Skylake\end{tabular} & 
        None
        \\

        ZIB &
        HLRN-IV &
        \checkmark & \checkmark  &  & \checkmark &
        Intel Cascade Lake AP & 
        \hspace{-0.6em}\begin{tabular}[t]{l}NVIDIA A100,\\Intel PVC\end{tabular} 
        \\

        \midrule
        North America & & & & & & & \\
        \midrule

        \hspace{-0.6em}\begin{tabular}[t]{l}Carnegie Mellon\\College of\\Engineering\end{tabular}&
        \hspace{-0.6em}\begin{tabular}[t]{l}Arjuna,\\Hercules\end{tabular} &
        \checkmark & \checkmark & \checkmark & \checkmark &
        \hspace{-0.6em}\begin{tabular}[t]{l}Intel Xeon,\\AMD Milan\end{tabular} &
        \hspace{-0.6em}\begin{tabular}[t]{l}NVIDIA A100,\\NVIDIA K80\end{tabular}
        \\

        Dartmouth College &
        Discovery &
        \checkmark &  & \checkmark & \checkmark &
        \hspace{-0.6em}\begin{tabular}[t]{l}Various Intel Xeon,\\AMD Rome\end{tabular} &
        NVIDIA V100
        \\

        FARSC (Harvard) &
        Cannon &
        \checkmark &  & \checkmark & \checkmark &
        Intel Cascade Lake &
        \hspace{-0.6em}\begin{tabular}[t]{l}NVIDIA V100,\\NVIDIA A100\end{tabular}
        \\

        HPC \@ LLNL &
        Various Systems &
        \checkmark &  & \checkmark & \checkmark &
        Various Processors &
        Various GPUs
        \\
        
        OLCF &
        Frontier/Crusher &
        \checkmark & \checkmark & \checkmark &  &
        \hspace{-0.6em}\begin{tabular}[t]{l}AMD Epyc\end{tabular} &
        AMD MI250X
        \\

        NERSC &
        Cori &
        \checkmark & \checkmark & \checkmark & \checkmark &
        \hspace{-0.6em}\begin{tabular}[t]{l}Intel Haswell,\\Intel KNL,\\Intel Skylake\end{tabular} &
        NVIDIA V100
        \\

        NERSC &
        Perlmutter &
        \checkmark & \checkmark & \checkmark & \checkmark &
        AMD Milan &
        NVIDIA A100
        \\

        \hspace{-0.6em}\begin{tabular}[t]{l}Open Science Grid\end{tabular} &
        &
        \text{\sffamily X} & \checkmark &  & \checkmark &
        Various Processors &
        Various GPUs
        \\

        Perimeter Institute for Theoretical Physics &
        Symmetry &
        \checkmark & \checkmark & \checkmark & \text{\sffamily X} &
        \hspace{-0.6em}\begin{tabular}[t]{l}AMD Epyc,\\Intel Xeon,\end{tabular} &
        NVIDIA A100
        \\

        Pittsburgh Supercomputing Center &
        Bridges-2 &
        \checkmark & \checkmark & \checkmark & \checkmark &
        \hspace{-0.6em}\begin{tabular}[t]{l}AMD Epyc,\\Intel Xeon,\end{tabular} &
        NVIDIA V100
        \\

        Princeton University &
        Several (including Tiger) &
        \checkmark & \checkmark & \checkmark & \checkmark &
        \hspace{-0.6em}\begin{tabular}[t]{l}Intel Skylake,\\Intel Broadwell\end{tabular} &
        NVIDIA P100
        \\

        \bottomrule
    \end{tabular}
    \vspace*{1em}
\caption{%
    August 8, 2022 snapshot of the Julia support level at different HPC centers (current list is available at \url{https://github.com/hlrs-tasc/julia-on-hpc-systems}). User support legend: P = official version preinstalled, U = center provides user support (e.g., center staff answers user questions), I = support for interactive workflows, and D = center provides documentation.%
    \label{tab:center-support}%
}
\end{table*}

Current support at Oak Ridge National Laboratory's Oak Ridge Leadership Computing Facility (OLCF)~\citep{OLCF} for Summit and Crusher, which is Frontier's test bed system, include recent Julia versions in the user modules. Similarly, the OLCF JupyterHub interface provides custom multithreaded Julia kernels for access to the high-performance file systems. Although user support is available, gaps exist in the official documentation and training~\citep{9167301}, and these gaps must be closed to make Julia a viable option for exascale computing.

\section{Teaching}
\label{sec:teaching}

Julia's dynamic characteristics and interactive features make it a powerful entry-level tool for teaching, and the official Julia website\footnote{\url{https://julialang.org/learning/classes/}} offers a selection of online courses. Examples include the Massachusetts Institute of Technology (MIT) modern numerical computing course using Julia for a decade\footnote{\url{http://courses.csail.mit.edu/18.337/2018}}. While ETH Zurich offers a GPU for HPC programming classes using Julia\footnote{\url{https://pde-on-gpu.vaw.ethz.ch}}. The high-level of abstraction enables classroom experiences comparable to Python or MATLAB, and the rich collection of scientific libraries spans a broad spectrum of applications. As an answer to the two-language problem, Julia can empower domain scientists to dive into HPC development, thereby removing most of the usual barriers that the endeavor would encounter. As such, Julia offers a fast track for domain scientists interested in promoting the development of code on a high level while also offering opportunities for further optimizations, performance engineering, and native tools for precise code analysis.

\subsection{Code introspection and performance engineering} 
\label{subsec:high-low}


In addition to Julia's REPL (read-eval-print loop) component, interactive interfaces such as Jupyter\footnote{Although Jupyter supports several languages, it derives its name from three programming languages: Julia, Python, and R.}~\citep{jupyter} and Pluto~\citep{pluto} provide an engaging learning environment for students with a low barrier to entry. Combined with Julia's high-level syntax, readily available 2D and 3D visualization packages such \texttt{Plots.jl}~\citep{plots_jl} and \texttt{Makie.jl}~\citep{DanischKrumbiegel2021}, and a built-in package manager---which also reliably delivers binary dependencies across different operating systems---these frameworks allow one to dive right into the concepts of interest rather than dealing with distracting technicalities or working around missing language features.

At the same time, Julia's just-ahead-of-time compilation delivers fast and pure native code by leveraging the modular LLVM compiler infrastructure. This distinguishes Julia from other dynamic high-level languages, which are typically several orders of magnitude slower, and puts it in the ranks of traditional HPC programming languages (e.g., C, Fortran) in terms of performance and low-level interpretability. As for the latter, the built-in introspection tools, \jlinl{@code\_typed}, \jlinl{@code\_llvm}, and \jlinl{@code\_native}, provide a unique way to interactively explore the compilation of high-level Julia code to intermediate LLVM-IR and low-level machine instructions. In particular, this feature allows one to demonstrate the connection between different variants of code and their respective performance (e.g., owing to the presence or absence of Single Instruction Multiple Data [SIMD] vectorization).
Given Julia's competitive speed, students can readily use the  language's interactive capabilities to write, analyze, and improve their own domain-specific production codes, thereby making the effort of learning Julia much more profitable for their science.

\subsection{Transferable knowledge and experience}
\label{subsec:knowledge-transfer}
Teaching may become a challenging endeavour because it requires the instructor to extract the key concepts from a complex workflow and expose them to students as clear, simple, and concise incremental steps. Conciseness is crucial there because reducing complexity and new concepts to the strict minimum usually accounts for enhanced focus, which in turn enables a steeper learning curve. Teaching is mostly about introducing, exemplifying, and exercising new concepts. Julia's conciseness, performance, and interactive features enable the instructor to go through all these steps with a single code. Julia's high-level syntax permits the instructor to efficiently prototype new concepts into code, and that code actually executes with optimal performance. This is important when teaching algorithmic concepts because users/students usually do not like to wait for their algorithm to complete.

However, the story is dramatically different for HPC. In HPC, one would ideally have some simple high-level code snippets that demonstrate performance-oriented, often parallel and accelerator-based implementations with strong focus on run-time (or implementation) performance. 
High-level or interpreted languages will mostly fail at this stage because the algorithm design will remain conceptual or require a low-level implementation to fulfil the performance expectations, thereby introducing a significant barrier in the teaching workflow owing to the inherent complexity overhead. 
The same challenges apply when targeting accelerators such as GPUs. It may be possible to conceptually design GPU kernels in any language; however, when it comes to testing the actual implementation in terms of performance, one would obviously need to have a GPU-compatible code. Julia overcomes the two-language barrier as it allows a single high-level and concise code to be regrouped as the essence of the algorithm or implementation of interest and will most likely enable a high-performance execution of it---be it for demonstration or production purposes. The SAXPY code (\Cref{code:saxpy}) exemplifies this by achieving a memory throughput of $\sim$1,260~GB/s for a high-level broadcasting implementation and $\sim$1,350~GB/s for compact CUDA kernel and CUBLAS variants on an NVIDIA A100 SXM4 GPU.

Ultimately, students and users can learn about and experiment with basic and advanced HPC concepts within the same interactive language in a portable way. Teaching material can be prototyped on personal computers or laptops, and the same codes can be later deployed on GPUs or HPC servers without code duplication or explicit porting between languages. Moreover, Julia provides a single language to enable experimenting with HPC that can be readily deployed in domain sciences.

\begin{figure}
\begin{jllisting}
using CUDA
const dim = 100_000_000
const a = 3.1415

x = CUDA.ones(dim)
y = CUDA.ones(dim)
z = CUDA.zeros(dim)

# (a) SAXPY via high-level broadcasting
CUDA.@sync z .= a .* x .+ y

# (b) SAXPY via CUBLAS
CUDA.@sync CUBLAS.axpy!(dim, a, x, y)

# (c) SAXPY via CUDA kernel
function saxpy_gpu_kernel!(z, a, x, y)
    i = (blockIdx().x - 1) * blockDim().x +
        threadIdx().x
    if i <= length(z)
        @inbounds z[i] = a * x[i] + y[i]
    end
    return nothing
end

# launch configuration
nthreads = 1024
nblocks = cld(dim, nthreads)

# execute the kernel
CUDA.@sync @cuda(
    threads = nthreads,
    blocks = nblocks,
    saxpy_gpu_kernel!(z, a, x, y)
)
\end{jllisting}
\caption{Three different SAXPY implementations based on \texttt{CUDA.jl}~\citep{besard2018effective} for NVIDIA GPUs: (a)~high-level variant that utilizes broadcasting and array abstractions, (b)~simple call into the cuBLAS vendor library, and (c)~custom SAXPY CUDA kernel written in and launched from Julia.}
\label{code:saxpy}
\end{figure}

\section{Scalability and portability}
\label{sec:scalability-portability}
The ability to efficiently deploy a single HPC code on different architectures and at different scales is a key feature for productivity in scientific HPC. Julia offers features that help reduce the complexity of this task, including multiple dispatch, cost-less, high-level abstractions and extensive metaprogramming capabilities. As a result, powerful low- and high-level packages for performance-portable shared and distributed parallelization have emerged.

\subsection{Performance scalability}
\label{subsec:scalability}

Julia's base multithreading support and generic high-level packages (e.g., \texttt{LoopVectorization.jl}~\citep{loopvectorization_jl}, \texttt{SIMD.jl}~\citep{simd_jl}) enable straightforward intranode CPU parallelization. Packages such as \texttt{CUDA.jl}~\citep{besard2019backends}, \texttt{AMDGPU.jl}~\citep{amdgpu_jl}, and \texttt{OneAPI.jl}~\citep{oneapi_jl} provide the ability to run Julia code natively on GPUs.
Various domain- and method-specific packages (e.g., \texttt{ParallelStencil.jl}~\citep{parallelstencil_jl}, \texttt{Flux.jl}~\citep{flux_jl_2018, innes2018}) simplify efficient shared-memory parallelization on GPUs and CPUs for the targeted applications and make it accessible to domain scientists.

Julia includes a generic approach to distributed computing via the \texttt{Distributed.jl} module. A convenient and zero-overhead wrapper for MPI is also available via the \texttt{MPI.jl} package~\citep{byrne2021mpi}. \texttt{MPI.jl} supports CUDA- and ROCm-aware MPI and enables packages that build on it to leverage remote direct memory access (RDMA). Similarly, \texttt{MPI.jl} enables wrappers for MPI-based libraries for scalable parallel I/O
such as: \texttt{HDF5.jl}\footnote{\url{https://github.com/JuliaIO/HDF5.jl}} ~\citep{osti_1398484,hdf5}, and the more streaming oriented \texttt{ADIOS2.jl}\footnote{\url{https://github.com/eschnett/ADIOS2.jl}}~\citep{GODOY2020100561} for data storage and streaming at scale.
As for shared memory parallelization, high-level packages can render distributed parallelization simple and efficient for certain classes of applications. Examples include \texttt{ImplicitGlobalGrid.jl}~\citep{implicitglobalgrid_jl}, which builds on \texttt{MPI.jl} and renders efficient RDMA-enabled distributed parallelization of stencil-based GPU and CPU applications on a regular staggered grid almost trivial, and \texttt{DistributedArrays.jl}~\citep{distributedarrays_jl}, which is a global-array interface that relies on the \texttt{Distributed.jl} module.
 
\begin{figure}
    \centering
    \includegraphics[width=\linewidth]{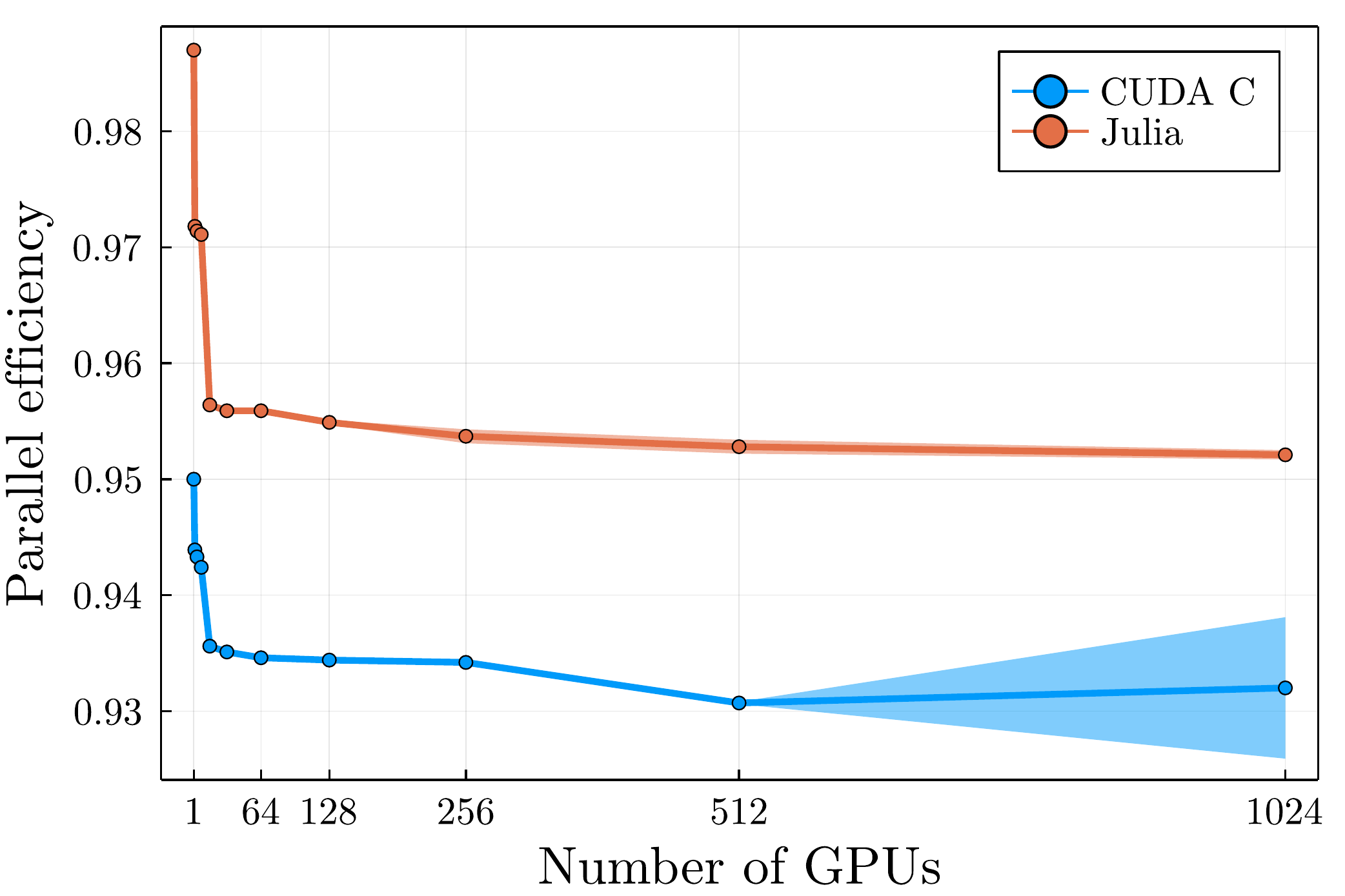}
    \caption{Parallel efficiency of a weak-scaling benchmark using 1 to 1,024 NVIDIA P100 GPUs on the Piz Daint Cray XC50. The blue and orange surfaces visualize the 95\% confidence interval of the reported medians. Adapted from~\citet{juliacon2020scaling}. The raw data and plotting script are available in the reproducibility repository~\citep{churavy2022bridgingRepro}.}
    \label{fig:par_eff}
\end{figure}

By combining high-level Julia packages for shared and distributed computing (e.g., \texttt{ParallelStencil.jl}, \texttt{ImplicitGlobalGrid.jl}), a single high-level HPC code can be readily deployed on a single CPU core or on thousands of CPUs or GPUs. The weak scaling of a Julia-based, coupled, hydro-mechanical 3D multiphysics solver achieves a parallel efficiency of more than 95\% on 1--1,024 NVIDIA Tesla P100 GPUs on the Piz Daint Cray XC50 supercomputer at the Swiss National Supercomputing Centre (\Cref{fig:par_eff}, adapted from~\citet{juliacon2020scaling}). These results were confirmed recently by close-to-ideal weak scaling achievements on up to 2,197 P100 GPUs~\citep{rass2022assessing}. The solver was written in CUDA C using MPI (blue data) and translated to Julia (red data) by using \texttt{ParallelStencil.jl} and \texttt{ImplicitGlobalGrid.jl}. On a single node, the Julia solver achieved 90\% of the CUDA C solver's performance (after the initial direct translation) without extensive Julia language--specific optimizations. It should be noted that we apply a strict definition of parallel efficiency, in which the reference performance for one GPU is given by the best known serial implementation in CUDA C and Julia. As a result, the reported parallel efficiency for one GPU is below 100\%, and this accounts for the performance loss caused by splitting boundary and inner-point calculations to enable communication/computation overlap (see \citet{gtc2019} for details). This performance loss was more significant for the CUDA C experiments than for the Julia experiments because less-refined parameters were used for the definition of the computation splitting. Thus, the results obtained with CUDA C could certainly be improved by redoing the experiments with better-suited parameters. 

\subsection{Performance portability}
\label{subsec:portability}

Julia's performance portability story unfolds along several main threads. First, Julia is capable of retargeting the language at a low-level for diverse platforms and accelerators. Second, library writers can use Julia's capabilities to build powerful abstractions. Last but not least, a common array abstraction allows for high-level performance-portable codes.

At the core of Julia's infrastructure sits a flexible and extensible compiler design and a multiple-dispatch language feature that enables code specialization for a given run-time type.

\paragraph{Array abstractions.}

Julia provides powerful array abstractions~\citep{Bezanson2017-ca} that when combined with several implementations allow the user
to efficiently express concepts in linear algebra, access optimized implementations, and retarget their programs.
At the core of the Julia standard library lies a common super-type, \jlinl{AbstractArray\{T,N\}}, for arrays with element type \jlinl{T} and \jlinl{N} dimensions. Many subtypes exist:
the dense array type \jlinl{Array\{T,N\}} (the most commonly used storage type for arrays allocated on the CPU), 
\jlinl{Tridiagonal\{T\}},
\jlinl{Transpose\{T,<:AbstractArray\{T,N\}\}} (a behavioral wrapper that transforms \jlinl{A[i,j]} into \jlinl{A[j,i]}), 
\jlinl{SparseMatrixCSC\{T\}}, and \jlinl{CUDA.CuArray\{T,N\}} (for arrays on NVIDIA GPUs). 
The \texttt{LinearOperators.jl}~\citep{orban-siqueira-linearoperators-2020} and \texttt{LinearMaps.jl}~\citep{linearmaps_jl} packages also provide types that implement linear operators specified as functions without storing any elements (i.e., matrix shell).

All subtypes of \jlinl{AbstractArray\{T,N\}} implement an \jlinl{N}-dimensional array with element type \jlinl{T}. The way in which elements are stored, which elements are stored, and how the various operations (e.g., addition, multiplication, element access, iteration) are used is left to the implementation. 
Typically, code that uses arrays (e.g., vectors, matrices, tensors) does not choose a particular implementation but works with any array type. This leads to the same freedom that Kokkos provides---storage and iteration implementation details are decoupled from the algorithms that use these arrays (as much as possible). New hardware back ends for accelerators can be supported in a straightforward manner by implementing the appropriate array storage types, similar to \jlinl{CUArray}.

The user can apply high-level abstractions (e.g., \jlinl{map}, \jlinl{reduce}, \jlinl{mapreduce}, broadcasting) as well as linear algebra routines and other numerical computing operations (e.g., Fourier transforms) to solve scientific problems. For example, the code in \Cref{code:ml} implements a simple train loop for a neural network. Notably, to
execute this code on the GPU, the user does not need to change the code itself---the user only has to move the data to the GPU. One can achieve this by adding \jlinl{x = CuArray(x)}, \jlinl{y = CuArray(y)}, and \jlinl{w = CuArray(w)} before the loop.

\begin{figure}
\begin{jllisting}
using LinearAlgebra
loss(w,b,x,y) = sum(abs2, y - (w*x .+ b)) / size(y,2)
loss∇w(w, b, x, y) = ...
lossdb(w, b, x, y) = ...
function train(w, b, x, y ; lr=0.1)
   w -= lmul!(lr, loss∇w(w, b, x, y))
   b -= lr * lossdb(w, b, x, y)
   return w, b
end
n = 100; p = 10
x = randn(n, p)'
y = sum(x[1:5, :]; dims=1) .+ randn(n)' * 0.1
w = 0.0001 * randn(1,p)
b = 0.0
for i in 1:50
   w, b = train(w, b, x, y)
end
\end{jllisting}
\caption{A neural network training loop that uses Julia's linear algebra routines.}
\label{code:ml}
\end{figure}

These abstractions are all implemented in Julia itself. Most often, they are dispatched to optimized and specialized operations appropriate for the compute device as well as libraries that provide optimized BLAS operations.

Because the implementation is primarily in Julia, an enterprising user can provide a specialized array implementation and leverage the structure in their own problem. We demonstrate such an scenario in \Cref{code:hpc_array}. The user can create a wrapper array to encode mathematical knowledge into the array type.  In 
this case, the user needs $n$ numbers to represent a matrix that is dense but structured.
The user knows a special algorithm for the largest eigenvalue.  With the higher-level abstractions,
essentially the same code works on a single CPU, in a distributed setting, or on a GPU.

\begin{figure}
\begin{jllisting}
using LinearAlgebra

# Build a custom array type
struct DMatrix{T, V<:AbstractVector{T}} <: AbstractMatrix{T}
    v::V
end

Base.size(A::DMatrix) = length(A.v), length(A.v)
Base.getindex(A::DMatrix,i,j) = 
  A.v[i]*(i==j) + A.v[i]*A.v[j]

# Eigensolver for DMatrix
f(A::DMatrix) =
  λ -> 1 + mapreduce(v -> v^2 / (v - λ)  , +, A.v)
f′(A::DMatrix) =
  λ ->     mapreduce(v -> v^2 / (v - λ)^2, +, A.v)

import LinearAlgebra: eigmax
function eigmax(A::DMatrix; tol = eps(2.0))
    x₀ = maximum(A.v) + maximum(A.v)^2
    δ = f(A)(x₀) / f′(A)(x₀)
    while abs(δ) > x₀ * tol               
        x₀ -= δ
        δ = f(A)(x₀) / f′(A)(x₀) 
    end
    x0
end
\end{jllisting}
\caption{A user-defined array type that only stores a vector, $v$, yet presents the full matrix
$vv^T + \textrm{diag}(v)$
to indexing operations. A custom largest-eigenvalue-solver makes efficient use of this structure via multiple dispatch. Adapted from~\citet{Edelman2019-lr}.}
\label{code:hpc_array}
\end{figure}

\begin{figure}
\begin{jllisting}
using Distributed
addprocs(4)
using CUDA
using DistributedArrays

N = 4_000_000
v = randn(N)*0.1
A = DMatrix(v)

# Explicit data-movement
distA = DMatrix(distribute(v))
gpuA = DMatrix(CuArray(v))

# Execute eigmax on the CPU,
# distributed across multiple processes,
# and on a GPU.
eigmax(A)
eigmax(distA)
eigmax(gpuA)

\end{jllisting}
\caption{Transparent execution of a program in multiple execution domains.}
\label{code:hpc_array_usage}
\end{figure}

\paragraph{Powerful libraries.}

One guiding principle in Julia is that \emph{it is Julia all the way down}. Packages are implemented mostly in Julia itself, as is the base language, standard library, and parts of the compiler.
Consequently, there is very little \emph{special code}. By special code, we mean things that the base language (i.e., C or C\texttt{++}) can do that one could not instead implement 
in pure Julia as a package author. Because of this, there are very few cases in which users would need to write an extension in C or C\texttt{++}. 

That said, Julia does rely on external libraries to interact with the operating system and hardware, and it leverages these libraries when standard solutions already exist for common problems.

The combination of Julia's type system, compiler, efficient execution, metaprogramming and staged programming allows library authors to implement 
powerful libraries that interact with user code and other libraries. As an example, both \texttt{KernelAbstractions.jl} and \texttt{ParallelStencil.jl} use macros (metaprogramming) to extend the Julia language with new concepts.

The differential equation ecosystem uses higher-level functions and the
capability of the Julia compiler to specialize these higher-level functions
on the user-defined function, thereby leading to cross-optimization between the user and the
library code.

\paragraph{Compiling code.}

Starting at a function call, Julia selects and compiles the most specific function signature. First, Julia propagates the argument types through the body of the function by using an abstract interpretation. At this level, in-lining and constant propagation occur. Afterward, a few optimization passes written in Julia optimize the IR, and the optimized function is translated to LLVM-IR. Julia uses LLVM as a single-function optimizer and to perform scheduling optimization (e.g., loop-vectorization). Then, the function is emitted as a binary and linked in-memory using LLVM's ORC just-in-time.

\texttt{GPUCompiler.jl} reuses this infrastructure to collect all statically reachable functions into one LLVM module, which is then compiled and uploaded to the accelerators. This approach is shared among the packages that provide support for accelerators and is flexible enough to support new accelerators/compilation targets.

\texttt{GPUArrays.jl} provides generic abstractions and implementations of common functionalities on accelerators, and \texttt{KernelAbstractions.jl} provides an extension of the Julia language to write GPU kernels that can be retargeted to different accelerators.

\subsection{A language for both beginners and experts} 
Considerable resources must be invested to train a scientist or engineer to make effective use of HPC. This training typically starts with learning how to program in an undergraduate-level class that is not focused on HPC before being exposed to more advanced topics such as parallel programming, GPU programming, or performance optimization. Often, these introductory programming courses start with a language that is somewhat easy to learn, has a simple syntax, good support for interactivity and visualization, and a strong ecosystem with additional packages and learning material (e.g., Python or MATLAB).

However, this path can be problematic when users eventually switch to a high-performance language (e.g., C\texttt{++}, Fortran) to achieve the required performance for scientific or industrial projects that target compute clusters or supercomputers. As noted before, learning a new programming language is not trivial because concepts often do not translate one-to-one from one language to another, and oftentimes the new language's capabilities are not used to the fullest extent~\citep{ScholtzWiedenbeck90,ShresthaBottaEtAl20}.

The Julia programming language has the potential to overcome this division between easy-to-learn and fast-to-execute languages. Its simple base syntax allows novice programmers to quickly grasp basic concepts such as variables, control flow, or data structures with a convenient style that enables the translation of many mathematical formulae directly into code. Because it compiles to native code, Julia provides the efficiency and optimization opportunities required for production-type computations. This means that as users move to more advanced programming concepts and applications, they continuously accumulate and extend their experience with their programming language and do not need to switch between different tools for rapid prototyping or large-scale application programming. Because Julia provides a REPL, a compiler, and a package manager in one combined solution, it further eases the transition of users between their own laptops, a university cluster, or an extreme-scale machine. Tools, packages, and experience can seamlessly move between different systems and applications.

\subsection{Workflow portability and reusability}

As demonstrated by NERSC's Superfacility Project~\citep{bard2022lbnl}, HPC workloads are rapidly expanding beyond the boundaries of a single data center. At present, efforts to develop multisite workflows are driven by the increasing need to integrate HPC into the data analysis pipelines of large experiments. Furthermore, future DOE initiatives (e.g., the AI for science initiative~\citep{osti_1604756}) emphasize the need for cross-facility workflows. These developments are gradually shifting the emphasis from the HPC application, which must be tailored to specific hardware and software environments, to workflows that incorporate many applications and services at multiple data centers.

Previous studies of state-of-the-art cross-site workflows (e.g., \cite{antypas2021,giannakou2021}) provide a rough anatomy of cross-site workflows, which consist of (1) a data movement layer, (2) portable executables, (3) a workflow orchestration engine, and (4) a control layer that coordinates resources across facilities.

As described in \Cref{subsec:portability}, Julia's syntax provides a natural way to abstract away details of the system's hardware. This abstraction method is aided by the many packages that adopt \texttt{Preferences.jl},\footnote{\url{https://juliaparallel.org/tutorials/preferences/}} which allows HPC center administrators to configure site-specific settings (e.g., MPI). Notably, users do not need to follow a different deployment recipe for each site. Furthermore, the Julia HPC community is active in developing packages such as \texttt{MPItrampoline.jl} as well as bindings for Slurm and the Flux resource manager.

\section{Julia success stories}
\label{sec:julia success stories}

We have claimed that Julia is fast and useful for performance-critical
programs. This claim is backed up by the microbenchmarks on Julia's website\footnote{\url{https://julialang.org/benchmarks}, 
accessed 09-28-2021.} that show that Julia's performance is comparable to
compiled languages such as C and Fortran. Here, we corroborate this claim
with additional examples that range from low-level code to high-level libraries
and interfaces.

\subsection{Performance of the same algorithms}

\begin{figure}[ht]
    \centering
    \includegraphics[width=\linewidth]{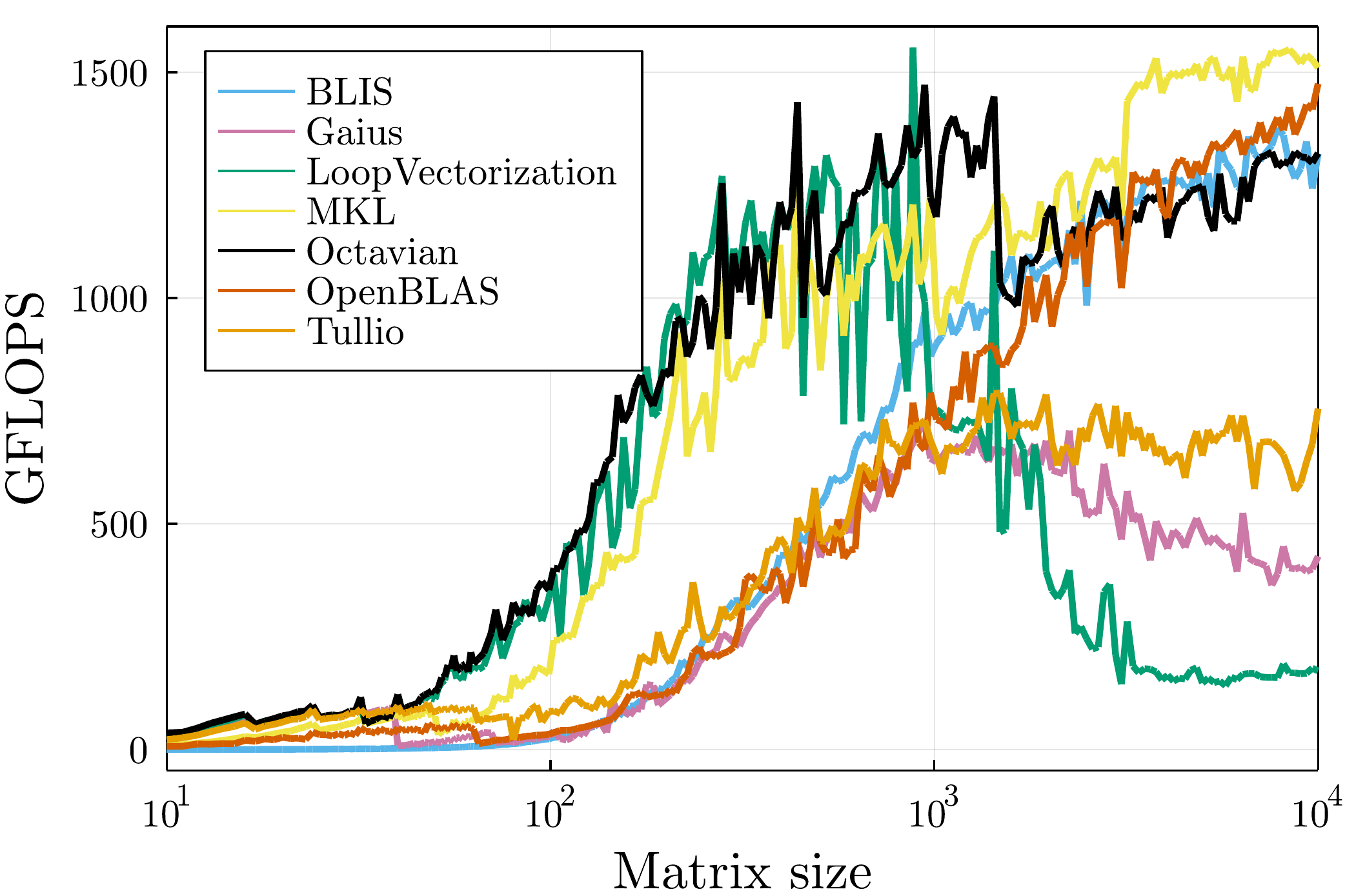}
    \caption{Benchmark of matrix multiplication using different BLAS libraries on a single Intel Xeon Gold Skylake 6148 CPU. The raw data and plotting script are available in the reproducibility repository~\citep{churavy2022bridgingRepro}. Inspired by a similar plot in \texttt{Octavian.jl}~\citep{octavian_jl}.}
    \label{fig:julia_blas}
\end{figure}

Julia can generate efficient
machine code for low-level BLAS routines (e.g., matrix multiplication),
which are used in various scientific workflows, including machine learning,
optimization, statistics, and numerical solution of differential equations.
\citet{elrod2021roadmap} demonstrated that highly optimized pure Julia packages (e.g., \texttt{Octavian.jl}) can be
on par with or even faster than established BLAS libraries (e.g., 
OpenBLAS, Intel MKL) on Intel's CPU hardware (\Cref{fig:julia_blas}). This is expected because Julia can generate similar LLVM-IR representations that could match the performance of the assembly code from these highly optimized libraries.

Similar results were obtained for discretizations of ordinary differential
equations, which are used in biology, chemistry, and pharmacology. Some example
benchmarks\footnote{\url{https://benchmarks.sciml.ai}, accessed 09-28-2021.}
that compare implementations of the same algorithm~\citep{dormand1980family}
in Fortran\footnote{\url{http://www.unige.ch/~hairer/software.html},
accessed 09-28-2021.} and Julia~\citep{rackauckas2017differentialequations}
show that the Julia versions are at least comparable to the Fortran codes and are
sometimes even more efficient owing to the enhanced in-lining and other
optimizations.
These results, which show a comparison of the same numerical methods implemented in different
programming languages, extend to partial differential equations,
hyperbolic conservation laws, and other transport-dominated phenomena
used in weather prediction, climate modeling, and aircraft design.
\citet{ranocha2021adaptive} compared the performance of the 
\texttt{Trixi.jl}~\citep{schlottkelakemper2021purely} Julia package with the mature
Fortran code FLUXO\footnote{\url{https://gitlab.com/project-fluxo/fluxo},
accessed 09-28-2021.} to implement the same algorithms for hyperbolic
conservation laws. The Julia code was at least as fast as the Fortran code and
sometimes up to 2$\times$ faster. More recently,~\citet{9652798} showed that in benchmarks across several HPC systems equipped with CPUs and GPUs, Julia's performance either matches or is only slightly behind existing parallel programming frameworks coded in C, C\texttt{++}, and Fortran.

\subsection{Algorithmic improvements}

Further evidence of Julia's performance and strengths is provided by
the \texttt{Gridap.jl} Julia package~\citep{badia2020gridap}, which can be used for
finite element discretizations in structural engineering, heat transfer
problems, and incompressible fluid flows. Leveraging Julia's
expressiveness and just-in-time compilation, \citet{verdugo2021software}
reported a finite element assembly performance comparable to FENICS~\citep{logg2010dolfin}, which is based on a DSL and
code generation via C/C\texttt{++}. Thus, Julia's expressiveness allows one to have
a code that is easier to develop and maintain without sacrificing
performance.
Furthermore, Julia makes it easier to develop new algorithms with direct
support for parallelism, thereby enabling significant speedups in applications
that benefit from algorithmic improvements (e.g., pharmaceutical
development\footnote{\url{https://juliacomputing.com/case-studies/pfizer},
accessed 09-28-2021.}).

\subsection{Common interfaces}

One of Julia's strengths is the use of
common interfaces in libraries enabled by multiple dispatch. For example, the standard array interface
is generic and allows the use of CPUs and GPUs~\citep{besard2019backends}.
Furthermore, automatic differentiation and other tasks do not rely on
creating a new array type; instead, they can reuse existing functionality.
By using generic programming based on these common interfaces in Julia, packages
can work together seamlessly without boilerplate glue code~\citep{karpinski2019unreasonable}.
For example, error propagation with \texttt{Measurements.jl} can be combined with
spatial semi-discretizations from \texttt{Trixi.jl} and time integration methods
from \texttt{OrdinaryDiffEq.jl} for numerical simulations without special glue code.
Additionally, the results can be visualized directly with \texttt{Plots.jl}.

At a lower level, common interfaces and operator overloading enable
automatic differentiation~\citep{revels2016forward}, speedups provided
by using low- and mixed-precision arithmetic on modern hardware~\citep{klower2020number},
and uncertainty propagation~\citep{giordano2016uncertainty}.

At a higher level, such common interfaces are useful for
algorithms in certain problem classes:
solving linear systems,\footnote{\url{https://github.com/SciML/LinearSolve.jl}, accessed 03-01-2022.}
differential equations~\citep{rackauckas2019confederated},
mathematical optimization~\citep{legat2020mathoptinterface},
and automatic differentiation~\citep{schafer2021abstractdifferentiation}. Because the optimal choice of a numerical
algorithm depends on the problem, providing all algorithms via a unified
interface enables users to swap algorithms depending on their needs.
There are focused research efforts to organize such
open interfaces to allow seamless interconnection in scientific computations (e.g., in the Mathematical Research Data Initiative\footnote{\url{https://www.mardi4nfdi.de}, accessed 03-01-2022.}).
\citet{dunning2017jump} demonstrated how such common interfaces can be
used via an open-source modeling language for optimization in Julia that
is competitive with widely used commercial systems and can even outperform
other open-source alternatives.

\subsection{Julia's adoption in CSE}

Given its features and performance, Julia has demonstrated
its readiness for the diverse set of applications in the broader CSE field.
Furthermore, we see this readiness as an opportunity for HPC.
Working well with CSE applications is crucial for the success of Julia in HPC because these applications allow for testing proven technologies and algorithms at different scales with varying levels of support in a broad community. Success stories in different CSE fields include
algebraic geometry~\citep{breiding2018homotopycontinuation},
astronomy at petascale~\citep{regier2018cataloging},
cancer therapies~\citep{pich2019mutational},
computer algebra and number theory~\citep{fieker2017nemo},
electrical engineering~\citep{plietzsch2022powerdynamics},
epidemic modeling~\citep{weitz2020modeling},
high-performance geophysical simulations~\citep{rass2022assessing},
fluid dynamics~\citep{ramadhan2020oceananigans,ranocha2021adaptive},
semiconductor theory~\citep{frost2017calculating},
symbolic-numeric computing~\citep{ketcheson2021computing,iravanian2022,ma2021modelingtoolkit},
quantum optics~\citep{kramer2018quantumoptics},
quantum chemistry~\citep{aroeira2022fermi},
quantum physics~\citep{DFTKjcon},
and many others.

Typically, the performance of these Julia packages is at least comparable to
existing frameworks in low-level programming languages. Sometimes Julia's
productivity features even enable improved algorithmic development and
simpler reuse of existing specialized implementations, thereby leading to speedups
compared to established codes. If highly tuned libraries of core routines
are already available with a C interface, then they can be easily accessed from Julia. 
Thus, a gradual transition that incorporates old code bases is
also feasible, as described in~\Cref{sec:interoperability}.

\section{Interoperability and composability with preexisting code}
\label{sec:interoperability}

Owing to the large investment in creating, optimizing, and maintaining HPC software infrastructure, developers do not have to throw away or rewrite their Fortran, C, or C\texttt{++} codes. Interoperability with preexisting codes has been a top priority and is at the heart of Julia's advantage. Furthermore, to be successful in this space, one must reuse the tremendous work from well-established HPC frameworks. Although there is interest in writing BLAS routines in pure Julia~\citep{elrod2021roadmap} (\Cref{fig:julia_blas}), the ability to call existing vendor-optimized BLAS libraries was important to kick-start the language ecosystem. In \Cref{subsec:pre-existing}, we describe how this capability has grown
to integrate preexisting HPC codes into Julia. \Cref{par:enzyme} describes how these codes can be enhanced with new capabilities.
Additionally, \Cref{subsec:calling-julia} describes how Julia can be used as an 
implementation language for new algorithms, thus requiring Julia to be embedded into preexisting HPC software.

\subsection{Calling existing codes from Julia}
\label{subsec:pre-existing}

HPC workflows are becoming increasingly complex as a result of increasing resource heterogeneity as well as a growing need for HPC in traditionally non-HPC domains. Yet, traditional HPC code bases are written in languages that prioritize bare-metal performance, and this focus results in low productivity when developing workflows. As a result, we need a programming language that can express complex workflows while still making use of existing codes that encapsulate a large amount (often decades) of institutional and domain knowledge. A common example is incorporating simulation codes and solvers into experimental data analysis workflows.

By far the most common approach in HPC has been to adopt Python as the workflow language and develop high-performance kernels in HPC languages. This approach has a problem: the workflow orchestration layer is not optimized for HPC.
\begin{table*}
    \center
    \begin{tabular}{lllllr}
    \toprule
        \textbf{Function Signature}    & \multicolumn{2}{c}{\textbf{Pybind11}} & \multicolumn{2}{c}{\textbf{Julia's {\tt ccall}}} & \textbf{Speedup}\\
        \midrule
        {\tt int fn0()}                     & 132 & $\pm 14.9$ & 2.34 & $\pm 1.24$  & $56 \times$  \\
        {\tt int fn1(int)}                  & 217 & $\pm 20.9$ & 2.35 & $\pm 1.33$  & $92 \times$  \\
        {\tt double fn2(int, double)}       & 232 & $\pm 11.7$ & 2.32 & $\pm 0.189$ & $100 \times$ \\
        {\tt char* fn3(int, double, char*)} & 267 & $\pm 28.9$ & 6.27 & $\pm 0.396$ & $42 \times$ \\
        \bottomrule
    \end{tabular}
    \vspace*{1em}
\caption{%
    Round-trip times for calling C functions from Python (using Pybind11) and Julia (using {\tt ccall}). All times are in nanoseconds. Round-trip times in Python include the time to resolve the function symbol, convert Python types to native C-types, invoke the function call, and return the result (including the conversion of the returned C-type to native Python types). Because C-types are binary-compatible with Julia data types, the Julia benchmark does not require type conversions. The benchmark results were collected by using an Intel Core i7-1185G7 CPU running at 3.00~GHz with Julia version 1.7.1, Python version 3.8.10, and Pybind11 version 2.9.1.
    All scripts required to reproduce these results are available in the reproducibility repository~\citep{churavy2022bridgingRepro}.%
    \label{tab:fn-calls}%
}
\end{table*}
To illustrate this problem, we compare the round-trip time to call a C function with Pybind11~\citep{pybind11} vs. Julia's native {\tt ccall} interface (see~\Cref{tab:fn-calls} for results). The need to convert between Python data types and native C data types can be seen as an increased round-trip time in the Pybind11 benchmark results. Therefore, workflows coordinated by using Python codes will avoid frequent calls to small C functions---instead opting to combine work in monolithic C kernels. Julia does not have this limitation.

\paragraph{Adding new capabilities to preexisting code.}
\label{par:enzyme}

Over the last few years, Julia has become a test bed for the development of new techniques in probabilistic programming~\citep{10.1145/3314221.3314642,ge2018t} as well as scientific machine learning~\citep{rackauckas2020sciml}. For these new
techniques, the availability of gradients through automatic differentiation has been key. Similarly, the CESMIX project at the MIT is currently building an integrated framework for uncertainty quantification that greatly benefits from the availability of gradients.

Although Julia has emphasized interoperability with codes written in C, C\texttt{++}, or Fortran from the very beginning, there is an open question as to whether these new techniques can be utilized in codes that are a mixture of Julia~+~$x$, where $x$ is an HPC application to which one wishes to apply these techniques. The lynchpin for any attempt at this will be the availability of gradients and the integration of those gradients into Julia's automatic-differentiation frameworks.

Enzyme~\citep{moses2020instead} and its \texttt{Enzyme.jl} Julia front end are an automatic differentiation framework that operates over the LLVM-IR (instead of operating in operator-overloading or source-rewriting modes) and can thus synthesize gradients for multiple languages as long as they have an LLVM front end. This means it supports C, C\texttt{++}, Julia, and Rust with experimental support for Fortran. Enzyme can be used for differentiating large C\texttt{++} projects as well as CUDA and HIP GPU kernels~\citep{Moses2021-ai}. Support for additional forms of parallelism (e.g., OpenMP, MPI) is part of the roadmap.

By leveraging Enzyme, users can perform cross-language automatic-differentiation and thus integrate newly developed capabilities in Julia with previously existing HPC libraries.

\subsection{Calling Julia from C}
\label{subsec:calling-julia}

Fully featured Julia HPC code can be compiled into C libraries and called from regular C applications, as shown in a proof of concept with a MultiGPU 2D heat diffusion solver written in Julia and using CUDA, MPI, and graphics called from C.\footnote{\url{https://github.com/omlins/libdiffusion}}

The proof of concept shows that variables can be passed from C to Julia in a straightforward and portable manner. The example passes a GPU array allocated and initialized in the C code and an MPI communicator created in the C code to the solver written in Julia. Furthermore, support of CUDA-aware MPI that leverages RDMA, which is frequently requested in HPC, was successfully demonstrated.

Straightforward scientific visualization is possible thanks to Julia's graphics packages. The proof of concept demonstrated this by producing an animated GIF using the \texttt{Plots.jl} package from within the generated C library. For additional productivity in scientific HPC code development, Julia code that is compiled to a C library (e.g., the heat diffusion solver in the proof of concept) can also be executed within the Julia run time in an interactive manner.

The library building is enabled by the \texttt{PackageCompiler.jl} julia package~\citep{packagecompiler_jl}.

\section{Now is the time for Julia in HPC}
\label{sec:conclusions}

We are seeing a rapid uptake of Julia in technical computing. Consequently, the interest in scaling up Julia applications for HPC and designing HPC applications in Julia from the start are also on the rise.

As with every new tool in HPC, the initial adoption must overcome challenges and to some extent adapt to the unique HPC environments. It is therefore encouraging that many HPC centers are already providing Julia to their users.

\subsection{For application developers}

The Julia language has reached a level of maturity and stability suitable for production code. Julia's language design features native performance tools, LLVM-based just-in-time compilation, and support for parallelism and hardware accelerators, and this support makes it convenient for developing high-performance applications.
Furthermore, Julia adopted many tools that enhance developer productivity, including tools for package management, code introspection, a powerful REPL, and a module system. This makes Julia one of the few high-productivity high-performance programming languages.

Historically, the adoption of programming languages in HPC has been driven by the popularity of software frameworks that are programmed in those languages. Therefore, as Julia-based frameworks rise in popularity, so will the Julia language. However, it is not necessary to wait for Julia's \textit{killer app} because HPC frameworks also have a long history of multilanguage development (e.g., calling Fortran functions from C, calling C functions from Python). Therefore, we encourage developers to begin incorporating Julia components within existing HPC frameworks with the added value of portable access to different hardware accelerator targets.

\subsection{For Julia language developers}

The Julia language is uniquely suited for high-productivity, high-performance code development because it already addresses many issues of developing HPC applications in other high-productivity languages. Therefore, the work for language developers is not insurmountable. At present, the adoption challenges described in this work mainly stem from HPC hardware being similar but still different from consumer-grade hardware. For example, many HPC file systems are not optimized for loading small files, thereby resulting in slower application startup times that contribute significantly to a job's overall wall time. Also, the software and networking environments are very different at HPC centers. Vendors often address this issue by requiring the code to be compiled with their compilers to ensure the use of system drivers---something that usually does not work out of the box and can require configuration.

The Julia community is already providing many solutions in this area and truly shines with a variety of successful and documented HPC use cases---including how deployment challenges were overcome. Julia language developers should therefore curate these use cases, incorporate solutions into the language standard (e.g., ahead-of-time compilation for demanding codes, global site configurations), and add useful examples to the Julia documentation. Finally, because the Julia language has reached a high level of maturity, the language developers should now begin to emphasize language stability.

\subsection{For HPC center operators}

One major adoption challenge we have encountered so far is the lack of vendor support in HPC. This was felt most acutely during the initial deployment of the OLCF's Summit supercomputer because Julia lacked support for IBM's PowerPC architecture. This is less of an issue now with architectures such as ARM's AArch64 being used in consumer devices, which provides more access and opportunity for the Julia open-source community to develop support for these architectures early on~\citep{giordano2022:a64fx}. HPC centers have a history of pioneering new architectures (e.g., RISC-V) and new accelerator designs, and it is important to collaborate with vendors to garner Julia support. This will obviously benefit Julia, but because Julia is based around the open-source LLVM project, it will also lead to a better open compiler ecosystem for HPC.

China's Sunway architecture is an interesting data point. \citet{Shang2022-pq} describes a variational quantum eigensolver written in Julia scaling up to 20 million cores. While details are sparse, we can determine that they ported Julia to the Sunway SW26010P architecture. Each SW26010P core is split into a management processing element (MPE) and 64 compute processing elements (CPEs). They developed support for running on both the MPE and CPE cores. The CPE cores are targeted in an offloading style by using the infrastructure built for Julia's general accelerator support.

\section*{Conclusion}
As described here, our view is that the Julia programming language provides an excellent investment opportunity for the HPC community. Julia's value proposition prioritizes the needs of HPC in the current era: programming models that closely align with science to make HPC accessible; a coordinated ecosystem approach for packaging, testing, code instrumentation, and interactive computing; a growing community;  a modern and pragmatic workflow composition strategy that interoperates with LLVM and existing HPC frameworks for simulation performance; and a powerful data science and AI unified ecosystem.
Not since Fortran has a programming language been designed specifically to target the needs of the broader scientific community. Julia incorporates modern software requirements into the language to enrich the end-to-end co-design process and lower the cost of the software development cycle---from idea to performance portability.
This is a pivotal time for the HPC community as it continues to march toward a more heterogeneous computing landscape in the post-Moore era, in which data-driven AI workflows become relevant for scientific discovery at scale. We believe that investing in the Julia language and enriching its ecosystem capabilities will pay dividends in easing current and future challenges associated with the increasing cost and complexity of multidisciplinary HPC endeavors.

\section*{Reproducibility}

The benchmarks shown in \Cref{tab:fn-calls} were run on NVIDIA P100 GPUs on the Swiss National Supercomputing Centre's Piz Daint Cray XC50 and are available in our reproducibility repository~\citep{churavy2022bridgingRepro}.
The BLAS benchmarks shown in \Cref{fig:julia_blas} were run on a single Intel Xeon Gold Skylake 6148 CPU in Noctua 1 at PC2 and are also available in our reproducibility repository~\citep{churavy2022bridgingRepro}.

\begin{acks}
VC and AE gratefully acknowledges funding from the National Science Foundation (OAC-1835443, OAC-2103804, AGS-1835860, and AGS-1835881) and DARPA under agreement number HR0011-20-9-0016 (PaPPa). This research was also made possible by the generosity of Eric and Wendy Schmidt by recommendation of the Schmidt Futures program, by the Paul G. Allen Family Foundation, Charles Trimble, and the Audi Environmental Foundation. This material is based upon work supported by the DOE's National Nuclear Security Administration under award number DE-NA0003965.

LR and SO acknowledge financial support from the Swiss University Conference and the Swiss Council of Federal Institutes of Technology through the Platform for Advanced Scientific Computing program. This work was supported by a grant from the Swiss National Supercomputing Centre under project ID c23 obtained via the PASC project GPU4GEO.

CB gratefully acknowledges the funding of this project by computing time provided by the
Paderborn Center for Parallel Computing (PC2). This work is partially funded by Paderborn University’s research award for ''GreenIT``, as well as the Federal Ministry of Education and
Research (BMBF) and the state of North Rhine-Westphalia as part of the NHR Program.

MSL gratefully acknowledges funding by the Deutsche Forschungsgemeinschaft (DFG, German
Research Foundation) project FOR-5409 (SNuBIC).

Parts of this research are supported by the Exascale Computing Project (17-SC-20-SC), a joint project of the DOE’s Office of Science and the National Nuclear Security Administration, responsible for delivering a capable exascale ecosystem, including software, applications, and hardware technology, to support the nation’s exascale computing imperative.  This research used resources of the Oak Ridge Leadership Computing Facility at the Oak Ridge National Laboratory, which is supported by the Office of Science of the U.S. Department of Energy under Contract No. DE-AC05-00OR22725. This research used resources of the National Energy Research Scientific Computing Center (NERSC), a U.S. Department of Energy Office of Science User Facility located at Lawrence Berkeley National Laboratory, operated under Contract No. DE-AC02-05CH11231.

Research at Perimeter Institute is supported in part by the Government of Canada through the Department of Innovation, Science and Economic Development and by the Province of Ontario through the Ministry of Colleges and Universities.

The views and opinions of authors expressed herein do not necessarily state or reflect those of the United States government or any agency thereof. The US government is authorized to reproduce and distribute reprints for government purposes notwithstanding any copyright notation herein.

Notice:  This manuscript has been authored by UT-Battelle LLC under contract DE-AC05-00OR22725 with DOE. The US government retains and the publisher, by accepting the article for publication, acknowledges that the US government retains a nonexclusive, paid-up, irrevocable, worldwide license to publish or reproduce the published form of this manuscript, or allow others to do so, for US government purposes. DOE will provide public access to these results of federally sponsored research in accordance with the DOE Public Access Plan (\url{http://energy.gov/downloads/doe-public-access-plan}).

\end{acks}

\bibliographystyle{SageH}
\bibliography{references}

\begin{thebibliography}{130}
\providecommand{\natexlab}[1]{#1}
\providecommand{\url}[1]{\texttt{#1}}
\providecommand{\urlprefix}{URL }
\expandafter\ifx\csname urlstyle\endcsname\relax
  \providecommand{\doi}[1]{DOI:\discretionary{}{}{}#1}\else
  \providecommand{\doi}{DOI:\discretionary{}{}{}\begingroup
  \urlstyle{rm}\Url}\fi

\bibitem[{Abadi et~al.(2015)Abadi, Agarwal, Barham, Brevdo, Chen, Citro,
  Corrado, Davis, Dean, Devin, Ghemawat, Goodfellow, Harp, Irving, Isard, Jia,
  Jozefowicz, Kaiser, Kudlur, Levenberg, Man\'{e}, Monga, Moore, Murray, Olah,
  Schuster, Shlens, Steiner, Sutskever, Talwar, Tucker, Vanhoucke, Vasudevan,
  Vi\'{e}gas, Vinyals, Warden, Wattenberg, Wicke, Yu and Zheng}]{tensorflow}
Abadi M, Agarwal A, Barham P, Brevdo E, Chen Z, Citro C, Corrado GS, Davis A,
  Dean J, Devin M, Ghemawat S, Goodfellow I, Harp A, Irving G, Isard M, Jia Y,
  Jozefowicz R, Kaiser L, Kudlur M, Levenberg J, Man\'{e} D, Monga R, Moore S,
  Murray D, Olah C, Schuster M, Shlens J, Steiner B, Sutskever I, Talwar K,
  Tucker P, Vanhoucke V, Vasudevan V, Vi\'{e}gas F, Vinyals O, Warden P,
  Wattenberg M, Wicke M, Yu Y and Zheng X (2015) {TensorFlow}: {L}arge-{S}cale
  {M}achine {L}earning on {H}eterogeneous {S}ystems.
\newblock \urlprefix\url{https://www.tensorflow.org/}.
\newblock Software available from tensorflow.org.

\bibitem[{Allen et~al.(2005)Allen, Chase, Hallett, Luchangco, Maessen, Ryu,
  Steele~Jr, Tobin-Hochstadt, Dias, Eastlund et~al.}]{allen2005fortress}
Allen E, Chase D, Hallett J, Luchangco V, Maessen JW, Ryu S, Steele~Jr GL,
  Tobin-Hochstadt S, Dias J, Eastlund C et~al. (2005) The {F}ortress language
  specification 139(140).

\bibitem[{Almasi(2011)}]{Almasi2011}
Almasi G (2011) \emph{PGAS (Partitioned Global Address Space) Languages}.
\newblock Boston, MA: Springer US.
\newblock ISBN 978-0-387-09766-4, pp. 1539--1545.
\newblock \doi{10.1007/978-0-387-09766-4_210}.
\newblock \urlprefix\url{https://doi.org/10.1007/978-0-387-09766-4_210}.

\bibitem[{AMD(2008)}]{rocm_hip}
AMD (2008) {ROCm} {HIP}: {H}eterogeneous-{C}omputing {I}nterface for
  {P}ortability.
\newblock \url{https://github.com/ROCm-Developer-Tools/HIP}.

\bibitem[{Antypas et~al.(2021)Antypas, Bard, Blaschke, Shane~Canon, Enders,
  Shankar, Somnath, Stansberry, Uram and Wilkinson}]{antypas2021}
Antypas KB, Bard DJ, Blaschke JP, Shane~Canon R, Enders B, Shankar MA, Somnath
  S, Stansberry D, Uram TD and Wilkinson SR (2021) Enabling discovery data
  science through cross-facility workflows.
\newblock In: \emph{2021 IEEE International Conference on Big Data (Big Data)}.
  pp. 3671--3680.
\newblock \doi{10.1109/BigData52589.2021.9671421}.

\bibitem[{Aroeira et~al.(2022)Aroeira, Davis, Turney and
  Schaefer~III}]{aroeira2022fermi}
Aroeira GJ, Davis MM, Turney JM and Schaefer~III HF (2022) Fermi.jl: {A}
  {M}odern {D}esign for {Q}uantum {C}hemistry.
\newblock \emph{Journal of Chemical Theory and Computation}
  \doi{10.1021/acs.jctc.1c00719}.

\bibitem[{Backus(1980)}]{backus1980programming}
Backus J (1980) {Programming in America in the 1950s—Some Personal
  Impressions}.
\newblock In: \emph{A History of Computing in the twentieth century}. Elsevier,
  pp. 125--135.

\bibitem[{Backus and Heising(1964)}]{4038201}
Backus JW and Heising WP (1964) Fortran.
\newblock \emph{IEEE Transactions on Electronic Computers} EC-13(4): 382--385.
\newblock \doi{10.1109/PGEC.1964.263818}.

\bibitem[{Badia and Verdugo(2020)}]{badia2020gridap}
Badia S and Verdugo F (2020) Gridap: {A}n extensible finite element toolbox in
  {J}ulia.
\newblock \emph{Journal of Open Source Software} 5(52): 2520.
\newblock \doi{10.21105/joss.02520}.

\bibitem[{Bard et~al.(2022)Bard, Snavely, Gerhardt, Lee, Totzke, Antypas,
  Arndt, Blaschke, Byna, Cheema, Cholia, Day, Enders, Gaur, Greiner, Groves,
  Kiran, Koziol, Lehman, Rowland, Samuel, Selvarajan, Sim, Skinner, Stephey,
  Thomas and Torok}]{bard2022lbnl}
Bard D, Snavely C, Gerhardt L, Lee J, Totzke B, Antypas K, Arndt W, Blaschke J,
  Byna S, Cheema R, Cholia S, Day M, Enders B, Gaur A, Greiner A, Groves T,
  Kiran M, Koziol Q, Lehman T, Rowland K, Samuel C, Selvarajan A, Sim A,
  Skinner D, Stephey L, Thomas R and Torok G (2022) The lbnl superfacility
  project report.

\bibitem[{Beckingsale et~al.(2019)Beckingsale, Burmark, Hornung, Jones,
  Killian, Kunen, Pearce, Robinson, Ryujin and Scogland}]{raja}
Beckingsale DA, Burmark J, Hornung R, Jones H, Killian W, Kunen AJ, Pearce O,
  Robinson P, Ryujin BS and Scogland TR (2019) {RAJA: Portable Performance for
  Large-Scale Scientific Applications}.
\newblock In: \emph{2019 IEEE/ACM International Workshop on Performance,
  Portability and Productivity in HPC (P3HPC)}. pp. 71--81.
\newblock \doi{10.1109/P3HPC49587.2019.00012}.

\bibitem[{Ben-Nun et~al.(2020)Ben-Nun, Gamblin, Hollman, Krishnan and
  Newburn}]{9309042}
Ben-Nun T, Gamblin T, Hollman DS, Krishnan H and Newburn CJ (2020) Workflows
  are the {N}ew {A}pplications: {C}hallenges in {P}erformance, {P}ortability,
  and {P}roductivity.
\newblock In: \emph{2020 IEEE/ACM International Workshop on Performance,
  Portability and Productivity in HPC (P3HPC)}. pp. 57--69.
\newblock \doi{10.1109/P3HPC51967.2020.00011}.

\bibitem[{Bercea et~al.(2016)Bercea, McRae, Ham, Mitchell, Rathgeber, Nardi,
  Luporini and Kelly}]{Firedrake}
Bercea GT, McRae ATT, Ham DA, Mitchell L, Rathgeber F, Nardi L, Luporini F and
  Kelly PHJ (2016) A structure-exploiting numbering algorithm for finite
  elements on extruded meshes, and its performance evaluation in {F}iredrake.
\newblock \emph{Geoscientific Model Development} 9(10): 3803--3815.
\newblock \doi{10.5194/gmd-9-3803-2016}.

\bibitem[{Besard et~al.(2018)Besard, Foket and De~Sutter}]{besard2018effective}
Besard T, Foket C and De~Sutter B (2018) Effective extensible programming:
  unleashing {J}ulia on {GPUs}.
\newblock \emph{IEEE Transactions on Parallel and Distributed Systems} 30(4):
  827--841.

\bibitem[{Besard et~al.(2019)Besard, Foket and De~Sutter}]{besard2019backends}
Besard T, Foket C and De~Sutter B (2019) Effective {E}xtensible {P}rogramming:
  {U}nleashing {J}ulia on {GPUs}.
\newblock \emph{IEEE Transactions on Parallel and Distributed Systems} 30(4):
  827--841.
\newblock \doi{10.1109/TPDS.2018.2872064}.

\bibitem[{Besard and other contributors(2020)}]{oneapi_jl}
Besard T and other contributors (2020) {oneAPI.jl}: {J}ulia support for the
  one{API} programming toolkit.
\newblock \url{https://github.com/JuliaGPU/oneAPI.jl}.

\bibitem[{Bezanson et~al.(2018)Bezanson, Chen, Chung, Karpinski, Shah, Vitek
  and Zoubritzky}]{bezanson2018julia}
Bezanson J, Chen J, Chung B, Karpinski S, Shah VB, Vitek J and Zoubritzky L
  (2018) Julia: {D}ynamism and performance reconciled by design.
\newblock \emph{Proceedings of the ACM on Programming Languages} 2(OOPSLA):
  1--23.
\newblock \doi{10.1145/3276490}.

\bibitem[{Bezanson et~al.(2017)Bezanson, Edelman, Karpinski and
  Shah}]{Bezanson2017-ca}
Bezanson J, Edelman A, Karpinski S and Shah VB (2017) Julia: A fresh approach
  to numerical computing.
\newblock \emph{SIAM Review} 59(1): 65--98.
\newblock \doi{10.1137/141000671}.

\bibitem[{Bradbury et~al.(2018)Bradbury, Frostig, Hawkins, Johnson, Leary,
  Maclaurin, Necula, Paszke, Vander{P}las, Wanderman-{M}ilne and Zhang}]{jax}
Bradbury J, Frostig R, Hawkins P, Johnson MJ, Leary C, Maclaurin D, Necula G,
  Paszke A, Vander{P}las J, Wanderman-{M}ilne S and Zhang Q (2018) {JAX}:
  composable transformations of {P}ython+{N}um{P}y programs.
\newblock \urlprefix\url{http://github.com/google/jax}.

\bibitem[{Breiding and Timme(2018)}]{breiding2018homotopycontinuation}
Breiding P and Timme S (2018) {HomotopyContinuation.jl}: A package for homotopy
  continuation in {J}ulia.
\newblock In: \emph{International Congress on Mathematical Software}. Springer,
  pp. 458--465.
\newblock \doi{10.1007/978-3-319-96418-8_54}.

\bibitem[{Buck(2007)}]{buck2007gpu}
Buck I (2007) {GPU} computing with {Nvidia} {CUDA}.
\newblock In: \emph{ACM SIGGRAPH 2007 courses}. pp. 6--es.

\bibitem[{Byna et~al.(2017)Byna, Chaarawi, Koziol, Mainzer and
  Willmore}]{osti_1398484}
Byna S, Chaarawi M, Koziol Q, Mainzer J and Willmore F (2017) Tuning hdf5
  subfiling performance on parallel file systems
  \urlprefix\url{https://www.osti.gov/biblio/1398484}.

\bibitem[{Byrne et~al.(2021)Byrne, Wilcox and Churavy}]{byrne2021mpi}
Byrne S, Wilcox LC and Churavy V (2021) {MPI}.jl: {J}ulia bindings for the
  {M}essage {P}assing {I}nterface.
\newblock In: \emph{Proceedings of the JuliaCon Conferences}, volume~1. p.~68.
\newblock \doi{10.21105/jcon.00068}.
\newblock \url{https://github.com/JuliaParallel/MPI.jl}.

\bibitem[{Carlsson and contributors(2022)}]{packagecompiler_jl}
Carlsson K and contributors (2022) {PackageCompiler.jl}: Compile your {J}ulia
  package.
\newblock \url{https://github.com/JuliaLang/PackageCompiler.j}.

\bibitem[{{Carter Edwards} et~al.(2014){Carter Edwards}, Trott and
  Sunderland}]{Kokkos}
{Carter Edwards} H, Trott CR and Sunderland D (2014) Kokkos: {E}nabling
  manycore performance portability through polymorphic memory access patterns.
\newblock \emph{Journal of Parallel and Distributed Computing} 74(12):
  3202--3216.
\newblock \doi{https://doi.org/10.1016/j.jpdc.2014.07.003}.
\newblock
  \urlprefix\url{https://www.sciencedirect.com/science/article/pii/S0743731514001257}.
\newblock Domain-Specific Languages and High-Level Frameworks for
  High-Performance Computing.

\bibitem[{Chamberlain et~al.(2007)Chamberlain, Callahan and
  Zima}]{doi:10.1177/1094342007078442}
Chamberlain B, Callahan D and Zima H (2007) Parallel {P}rogrammability and the
  {C}hapel {L}anguage.
\newblock \emph{The International Journal of High Performance Computing
  Applications} 21(3): 291--312.
\newblock \doi{10.1177/1094342007078442}.
\newblock \urlprefix\url{https://doi.org/10.1177/1094342007078442}.

\bibitem[{Christ et~al.(2022)Christ, Schwabeneder, Rackauckas, Borregaard and
  Breloff}]{plots_jl}
Christ S, Schwabeneder D, Rackauckas C, Borregaard MK and Breloff T (2022)
  Plots.jl -- a user extendable plotting api for the julia programming
  language.
\newblock \doi{10.48550/ARXIV.2204.08775}.
\newblock \urlprefix\url{https://arxiv.org/abs/2204.08775}.

\bibitem[{Churavy et~al.(2022)Churavy, Godoy, Bauer, Ranocha,
  Schlottke-Lakemper, R{\"a}ss, Blaschke, Giordano, Schnetter, Omlin, Vetter
  and Edelman}]{churavy2022bridgingRepro}
Churavy V, Godoy WF, Bauer C, Ranocha H, Schlottke-Lakemper M, R{\"a}ss L,
  Blaschke J, Giordano M, Schnetter E, Omlin S, Vetter JS and Edelman A (2022)
  Reproducibility repository for {B}ridging {HPC} communities through the
  {J}ulia programming language.
\newblock \url{https://github.com/JuliaParallel/paper-2022-HPC}.
\newblock \doi{10.5281/zenodo.7236016}.

\bibitem[{Contributors(2015)}]{distributedarrays_jl}
Contributors JP (2015) {DistributedArrays.jl}: {D}istributed {A}rrays in
  {J}ulia.
\newblock \url{https://github.com/JuliaParallel/DistributedArrays.jl}.

\bibitem[{Cusumano-Towner et~al.(2019)Cusumano-Towner, Saad, Lew and
  Mansinghka}]{10.1145/3314221.3314642}
Cusumano-Towner MF, Saad FA, Lew AK and Mansinghka VK (2019) Gen: A
  general-purpose probabilistic programming system with programmable inference.
\newblock In: \emph{Proceedings of the 40th ACM SIGPLAN Conference on
  Programming Language Design and Implementation}, PLDI 2019. New York, NY,
  USA: Association for Computing Machinery.
\newblock ISBN 9781450367127, p. 221–236.
\newblock \doi{10.1145/3314221.3314642}.
\newblock \urlprefix\url{https://doi.org/10.1145/3314221.3314642}.

\bibitem[{Dagum and Menon(1998)}]{openmp}
Dagum L and Menon R (1998) Open{MP}: an industry standard {API} for
  shared-memory programming.
\newblock \emph{IEEE computational science and engineering} 5(1): 46--55.

\bibitem[{Danisch and Krumbiegel(2021)}]{DanischKrumbiegel2021}
Danisch S and Krumbiegel J (2021) Makie.jl: Flexible high-performance data
  visualization for julia.
\newblock \emph{Journal of Open Source Software} 6(65): 3349.
\newblock \doi{10.21105/joss.03349}.
\newblock \urlprefix\url{https://doi.org/10.21105/joss.03349}.

\bibitem[{de~Graaf and contributors(2022)}]{pkgtemplates_jl}
de~Graaf C and contributors (2022) {PkgTemplates.jl}: {C}reate new {J}ulia
  packages, the easy way.
\newblock \url{https://github.com/invenia/PkgTemplates.jl}.

\bibitem[{Dongarra et~al.(2011)Dongarra, Beckman, Moore, Aerts, Aloisio, Andre,
  Barkai, Berthou, Boku, Braunschweig, Cappello, Chapman, Chi, Choudhary,
  Dosanjh, Dunning, Fiore, Geist, Gropp, Harrison, Hereld, Heroux, Hoisie,
  Hotta, Jin, Ishikawa, Johnson, Kale, Kenway, Keyes, Kramer, Labarta,
  Lichnewsky, Lippert, Lucas, Maccabe, Matsuoka, Messina, Michielse, Mohr,
  Mueller, Nagel, Nakashima, Papka, Reed, Sato, Seidel, Shalf, Skinner, Snir,
  Sterling, Stevens, Streitz, Sugar, Sumimoto, Tang, Taylor, Thakur, Trefethen,
  Valero, van~der Steen, Vetter, Williams, Wisniewski and
  Yelick}]{doi:10.1177/1094342010391989}
Dongarra J, Beckman P, Moore T, Aerts P, Aloisio G, Andre JC, Barkai D, Berthou
  JY, Boku T, Braunschweig B, Cappello F, Chapman B, Chi X, Choudhary A,
  Dosanjh S, Dunning T, Fiore S, Geist A, Gropp B, Harrison R, Hereld M, Heroux
  M, Hoisie A, Hotta K, Jin Z, Ishikawa Y, Johnson F, Kale S, Kenway R, Keyes
  D, Kramer B, Labarta J, Lichnewsky A, Lippert T, Lucas B, Maccabe B, Matsuoka
  S, Messina P, Michielse P, Mohr B, Mueller MS, Nagel WE, Nakashima H, Papka
  ME, Reed D, Sato M, Seidel E, Shalf J, Skinner D, Snir M, Sterling T, Stevens
  R, Streitz F, Sugar B, Sumimoto S, Tang W, Taylor J, Thakur R, Trefethen A,
  Valero M, van~der Steen A, Vetter J, Williams P, Wisniewski R and Yelick K
  (2011) The {I}nternational {E}xascale {S}oftware {P}roject roadmap.
\newblock \emph{The International Journal of High Performance Computing
  Applications} 25(1): 3--60.
\newblock \doi{10.1177/1094342010391989}.
\newblock \urlprefix\url{https://doi.org/10.1177/1094342010391989}.

\bibitem[{Dongarra et~al.(2008)Dongarra, Graybill, Harrod, Lucas, Lusk,
  Luszczek, McMahon, Snavely, Vetter, Yelick et~al.}]{dongarra2008darpa}
Dongarra J, Graybill R, Harrod W, Lucas R, Lusk E, Luszczek P, McMahon J,
  Snavely A, Vetter J, Yelick K et~al. (2008) {DARPA}'s {HPCS} program:
  {H}istory, models, tools, languages.
\newblock In: \emph{Advances in Computers}, volume~72. Elsevier, pp. 1--100.

\bibitem[{Dormand and Prince(1980)}]{dormand1980family}
Dormand JR and Prince PJ (1980) A family of embedded {R}unge-{K}utta formulae.
\newblock \emph{Journal of Computational and Applied Mathematics} 6(1): 19--26.
\newblock \doi{10.1016/0771-050X(80)90013-3}.

\bibitem[{Dunning et~al.(2017)Dunning, Huchette and Lubin}]{dunning2017jump}
Dunning I, Huchette J and Lubin M (2017) {JuM}p: {A} modeling language for
  mathematical optimization.
\newblock \emph{SIAM review} 59(2): 295--320.
\newblock \doi{10.1137/15M1020575}.

\bibitem[{Edelman(2019)}]{Edelman2019-lr}
Edelman A (2019) {IEEE-CS} sidney fernbach memorial award.
\newblock SC '19: Proceedings of the International Conference for High
  Performance Computing, Networking, Storage and Analysis.

\bibitem[{El-Ghazawi et~al.(2005)El-Ghazawi, Carlson, Sterling and
  Yelick}]{el2005upc}
El-Ghazawi T, Carlson W, Sterling T and Yelick K (2005) \emph{{UPC}:
  distributed shared memory programming}, volume~40.
\newblock John Wiley \& Sons.

\bibitem[{Elrod(2021)}]{elrod2021roadmap}
Elrod C (2021) Roadmap to {J}ulia {BLAS} and linear algebra.
\newblock \url{https://www.youtube.com/watch?v=KQ8nvlURX4M}. Talk presented at
  JuliaCon.

\bibitem[{Elrod et~al.(2022)Elrod, Aluthge, Protter and
  contributors}]{octavian_jl}
Elrod C, Aluthge D, Protter M and contributors (2022) Octavian.jl:
  {M}ulti-threaded {BLAS}-like library that provides pure {J}ulia matrix
  multiplication.
\newblock \url{https://github.com/JuliaLinearAlgebra/Octavian.jl}.

\bibitem[{Elrod and Lilly(2019)}]{loopvectorization_jl}
Elrod C and Lilly E (2019) {LoopVectorization.jl}: {M}acro(s) for vectorizing
  loops.
\newblock \url{https://github.com/JuliaSIMD/LoopVectorization.jl}.

\bibitem[{Enos et~al.(2010)Enos, Steffen, Fullop, Showerman, Shi, Esler,
  Kindratenko, Stone and Phillips}]{5598297}
Enos J, Steffen C, Fullop J, Showerman M, Shi G, Esler K, Kindratenko V, Stone
  JE and Phillips JC (2010) Quantifying the impact of {GPUs} on performance and
  energy efficiency in {HPC} clusters.
\newblock In: \emph{International Conference on Green Computing}. pp. 317--324.
\newblock \doi{10.1109/GREENCOMP.2010.5598297}.

\bibitem[{Fieker et~al.(2017)Fieker, Hart, Hofmann and
  Johansson}]{fieker2017nemo}
Fieker C, Hart W, Hofmann T and Johansson F (2017) Nemo/{H}ecke: computer
  algebra and number theory packages for the {J}ulia programming language.
\newblock In: \emph{Proceedings of the 2017 acm on international symposium on
  symbolic and algebraic computation}. pp. 157--164.
\newblock \doi{10.1145/3087604.3087611}.

\bibitem[{Frost(2017)}]{frost2017calculating}
Frost JM (2017) Calculating polaron mobility in halide perovskites.
\newblock \emph{Physical Review B} 96(19): 195202.
\newblock \doi{10.1103/PhysRevB.96.195202}.

\bibitem[{Ge et~al.(2018)Ge, Xu and Ghahramani}]{ge2018t}
Ge H, Xu K and Ghahramani Z (2018) Turing: a language for flexible
  probabilistic inference.
\newblock In: \emph{International Conference on Artificial Intelligence and
  Statistics, {AISTATS} 2018, 9-11 April 2018, Playa Blanca, Lanzarote, Canary
  Islands, Spain}. pp. 1682--1690.
\newblock \urlprefix\url{http://proceedings.mlr.press/v84/ge18b.html}.

\bibitem[{Giannakou et~al.(2021)Giannakou, Blaschke, Bard and
  Ramakrishnan}]{giannakou2021}
Giannakou A, Blaschke JP, Bard D and Ramakrishnan L (2021) Experiences with
  cross-facility real-time light source data analysis workflows.
\newblock In: \emph{2021 IEEE/ACM HPC for Urgent Decision Making (UrgentHPC)}.
  pp. 45--53.
\newblock \doi{10.1109/UrgentHPC54802.2021.00011}.

\bibitem[{Giordano(2016)}]{giordano2016uncertainty}
Giordano M (2016) Uncertainty propagation with functionally correlated
  quantities.

\bibitem[{{Giordano} et~al.(2022){Giordano}, {Kl{\"o}wer} and
  {Churavy}}]{giordano2022:a64fx}
{Giordano} M, {Kl{\"o}wer} M and {Churavy} V (2022) {Productivity meets
  Performance: Julia on A64FX}.
\newblock In: \emph{{2022 IEEE International Conference on Cluster Computing
  (CLUSTER)}}. pp. 549--555.
\newblock \doi{10.1109/CLUSTER51413.2022.00072}.

\bibitem[{Godoy et~al.(2020)Godoy, Podhorszki, Wang, Atkins, Eisenhauer, Gu,
  Davis, Choi, Germaschewski, Huck, Huebl, Kim, Kress, Kurc, Liu, Logan, Mehta,
  Ostrouchov, Parashar, Poeschel, Pugmire, Suchyta, Takahashi, Thompson,
  Tsutsumi, Wan, Wolf, Wu and Klasky}]{GODOY2020100561}
Godoy WF, Podhorszki N, Wang R, Atkins C, Eisenhauer G, Gu J, Davis P, Choi J,
  Germaschewski K, Huck K, Huebl A, Kim M, Kress J, Kurc T, Liu Q, Logan J,
  Mehta K, Ostrouchov G, Parashar M, Poeschel F, Pugmire D, Suchyta E,
  Takahashi K, Thompson N, Tsutsumi S, Wan L, Wolf M, Wu K and Klasky S (2020)
  Adios 2: The adaptable input output system. a framework for high-performance
  data management.
\newblock \emph{SoftwareX} 12: 100561.
\newblock \doi{https://doi.org/10.1016/j.softx.2020.100561}.

\bibitem[{Gropp et~al.(1999)Gropp, Gropp, Lusk, Skjellum and
  Lusk}]{gropp1999using}
Gropp W, Gropp WD, Lusk E, Skjellum A and Lusk ADFEE (1999) \emph{Using {MPI}:
  portable parallel programming with the message-passing interface}, volume~1.
\newblock MIT press.

\bibitem[{Hanson and Giordano(2021)}]{hanson:general}
Hanson EP and Giordano M (2021) Code, docs, and tests: what's in the {G}eneral
  registry?
\newblock \urlprefix\url{https://julialang.org/blog/2021/08/general-survey/}.

\bibitem[{Herbst et~al.(2021)Herbst, Levitt and Cancès}]{DFTKjcon}
Herbst MF, Levitt A and Cancès E (2021) {DFTK}: A {J}ulian approach for
  simulating electrons in solids.
\newblock \emph{Proc. JuliaCon Conf.} 3: 69.
\newblock \doi{10.21105/jcon.00069}.

\bibitem[{Heroux(2019)}]{heroux2019extreme}
Heroux MA (2019) The {E}xtreme-{S}cale {S}cientific {S}oftware {S}tack (e4s).
\newblock Technical report, Sandia National Lab.(SNL-NM), Albuquerque, NM
  (United States).

\bibitem[{Heroux et~al.(2018)Heroux, Carter, Thakur, Vetter, McInnes, Ahrens
  and Neely}]{osti_1463232}
Heroux MA, Carter J, Thakur R, Vetter J, McInnes LC, Ahrens J and Neely JR
  (2018) {ECP} {S}oftware {T}echnology {C}apability {A}ssessment {R}eport
  \doi{10.2172/1463232}.
\newblock \urlprefix\url{https://www.osti.gov/biblio/1463232}.

\bibitem[{HPCWire(2017)}]{hpcwire:celeste}
HPCWire (2017) Julia {J}oins {P}etaflop {C}lub.
\newblock
  \urlprefix\url{https://www.hpcwire.com/off-the-wire/julia-joins-petaflop-club/}.

\bibitem[{Innes(2018)}]{innes2018}
Innes M (2018) Flux: Elegant machine learning with julia.
\newblock \emph{Journal of Open Source Software} \doi{10.21105/joss.00602}.

\bibitem[{Innes et~al.(2018)Innes, Saba, Fischer, Gandhi, Rudilosso, Joy,
  Karmali, Pal and Shah}]{flux_jl_2018}
Innes M, Saba E, Fischer K, Gandhi D, Rudilosso MC, Joy NM, Karmali T, Pal A
  and Shah V (2018) Fashionable modelling with flux.
\newblock \emph{CoRR} abs/1811.01457.
\newblock \urlprefix\url{https://arxiv.org/abs/1811.01457}.

\bibitem[{Iravanian et~al.(2022)Iravanian, Martensen, Cheli, Gowda, Jain,
  {Julia Computing}, Ma and Rackauckas}]{iravanian2022}
Iravanian S, Martensen CJ, Cheli A, Gowda S, Jain A, {Julia Computing}, Ma Y
  and Rackauckas C (2022) Symbolic-numeric integration of univariate
  expressions based on sparse regression.

\bibitem[{Jakob et~al.(2017)Jakob, Rhinelander and Moldovan}]{pybind11}
Jakob W, Rhinelander J and Moldovan D (2017) pybind11 -- seamless operability
  between c++11 and python.
\newblock Https://github.com/pybind/pybind11.

\bibitem[{Janssens(2022)}]{cxxwrap_jl}
Janssens B (2022) {CxxWrap.jl}: {P}ackage to make {C\texttt{++}} libraries
  available in {J}ulia.
\newblock \url{https://github.com/JuliaInterop/CxxWrap.jl}.

\bibitem[{Johnson and contributors(2022)}]{pycall_jl}
Johnson SG and contributors (2022) {PyCall.jl}: {P}ackage to call {P}ython
  functions from the {J}ulia language.
\newblock \url{https://github.com/JuliaPy/PyCall.jl}.

\bibitem[{{Jupyter Development Team}(2022)}]{jupyter}
{Jupyter Development Team} (2022) \emph{{Jupyter}: Free software, open
  standards, and web services for interactive computing across all programming
  languages}.
\newblock \urlprefix\url{https://jupyter.org/}.

\bibitem[{Karpinski(2019)}]{karpinski2019unreasonable}
Karpinski S (2019) The {U}nreasonable {E}ffectiveness of {M}ultiple {D}ispatch.
\newblock \url{https://youtu.be/kc9HwsxE1OY}. Talk presented at JuliaCon.

\bibitem[{Karrasch et~al.(2022)Karrasch, Haegeman and
  contributors}]{linearmaps_jl}
Karrasch D, Haegeman J and contributors (2022) {LinearMaps.jl}: A {J}ulia
  package for defining and working with linear maps, also known as linear
  transformations or linear operators acting on vectors. {T}he only requirement
  for a {LinearMap} is that it can act on a vector (by multiplication)
  efficiently.
\newblock \url{https://github.com/JuliaLinearAlgebra/LinearMaps.jl}.

\bibitem[{Kedward et~al.(2022)Kedward, Aradi, Čertík, Curcic, Ehlert, Engel,
  Goswami, Hirsch, Lozada-Blanco, Magnin, Markus, Pagone, Pribec, Richardson,
  Snyder, Urban and Vandenplas}]{9736688}
Kedward LJ, Aradi B, Čertík O, Curcic M, Ehlert S, Engel P, Goswami R, Hirsch
  M, Lozada-Blanco A, Magnin V, Markus A, Pagone E, Pribec I, Richardson B,
  Snyder H, Urban J and Vandenplas J (2022) The {S}tate of {F}ortran.
\newblock \emph{Computing in Science \& Engineering} 24(2): 63--72.
\newblock \doi{10.1109/MCSE.2022.3159862}.

\bibitem[{Kernighan and Ritchie(1988)}]{kernighan1988c}
Kernighan BW and Ritchie DM (1988) \emph{The {C} programming language}.
\newblock Pearson Educaci{\'o}n.

\bibitem[{Ketcheson and Ranocha(2021)}]{ketcheson2021computing}
Ketcheson DI and Ranocha H (2021) Computing with {B}-series.

\bibitem[{Kindratenko et~al.(2009)Kindratenko, Enos, Shi, Showerman, Arnold,
  Stone, Phillips and Hwu}]{5289128}
Kindratenko VV, Enos JJ, Shi G, Showerman MT, Arnold GW, Stone JE, Phillips JC
  and Hwu Wm (2009) {GPU} clusters for high-performance computing.
\newblock In: \emph{2009 IEEE International Conference on Cluster Computing and
  Workshops}. pp. 1--8.
\newblock \doi{10.1109/CLUSTR.2009.5289128}.

\bibitem[{Kl{\"o}wer et~al.(2020)Kl{\"o}wer, D{\"u}ben and
  Palmer}]{klower2020number}
Kl{\"o}wer M, D{\"u}ben P and Palmer T (2020) Number formats, error mitigation,
  and scope for 16-bit arithmetics in weather and climate modeling analyzed
  with a shallow water model.
\newblock \emph{Journal of Advances in Modeling Earth Systems} 12(10):
  e2020MS002246.
\newblock \doi{10.1029/2020MS002246}.

\bibitem[{Kr{\"a}mer et~al.(2018)Kr{\"a}mer, Plankensteiner, Ostermann and
  Ritsch}]{kramer2018quantumoptics}
Kr{\"a}mer S, Plankensteiner D, Ostermann L and Ritsch H (2018)
  {QuantumOptics.jl}: A {J}ulia framework for simulating open quantum systems.
\newblock \emph{Computer Physics Communications} 227: 109--116.
\newblock \doi{10.1016/j.cpc.2018.02.004}.

\bibitem[{Lai and contributors(2022)}]{rcall_jl}
Lai R and contributors (2022) {RCall.jl}: {C}all {R} from {J}ulia.
\newblock \url{https://github.com/JuliaInterop/RCall.jl}.

\bibitem[{Lattner and Adve(2004)}]{1281665}
Lattner C and Adve V (2004) {LLVM}: a compilation framework for lifelong
  program analysis amp; transformation.
\newblock In: \emph{International Symposium on Code Generation and
  Optimization, 2004. CGO 2004.} pp. 75--86.
\newblock \doi{10.1109/CGO.2004.1281665}.

\bibitem[{Lavrijsen and Dutta(2016)}]{7836841}
Lavrijsen WT and Dutta A (2016) High-performance python-c++ bindings with pypy
  and cling.
\newblock In: \emph{2016 6th Workshop on Python for High-Performance and
  Scientific Computing (PyHPC)}. pp. 27--35.
\newblock \doi{10.1109/PyHPC.2016.008}.

\bibitem[{Legat et~al.(2020)Legat, Dowson, Garcia and
  Lubin}]{legat2020mathoptinterface}
Legat B, Dowson O, Garcia JD and Lubin M (2020) {MathOptInterface}: a data
  structure for mathematical optimization problems.

\bibitem[{Lin and McIntosh-Smith(2021)}]{9652798}
Lin WC and McIntosh-Smith S (2021) Comparing {J}ulia to {P}erformance
  {P}ortable {P}arallel {P}rogramming {M}odels for {HPC}.
\newblock In: \emph{2021 International Workshop on Performance Modeling,
  Benchmarking and Simulation of High Performance Computer Systems (PMBS)}. pp.
  94--105.
\newblock \doi{10.1109/PMBS54543.2021.00016}.

\bibitem[{Logg and Wells(2010)}]{logg2010dolfin}
Logg A and Wells GN (2010) {DOLFIN}: Automated finite element computing.
\newblock \emph{ACM Transactions on Mathematical Software (TOMS)} 37(2): 1--28.
\newblock \doi{10.1145/1731022.1731030}.

\bibitem[{Ma et~al.(2021)Ma, Gowda, Anantharaman, Laughman, Shah and
  Rackauckas}]{ma2021modelingtoolkit}
Ma Y, Gowda S, Anantharaman R, Laughman C, Shah V and Rackauckas C (2021)
  Modelingtoolkit: A composable graph transformation system for equation-based
  modeling.

\bibitem[{Marques and Barker(2020)}]{9167301}
Marques O and Barker A (2020) {T}raining {E}fforts in the {E}xascale
  {C}omputing {P}roject.
\newblock \emph{Computing in Science \& Engineering} 22(5): 103--107.
\newblock \doi{10.1109/MCSE.2020.3010596}.

\bibitem[{Mohamad and contributors(2022)}]{matlab_jl}
Mohamad M and contributors (2022) {MATLAB.jl}: {C}alling {MATLAB} in {J}ulia
  through {MATLAB} {E}ngine.
\newblock \url{https://github.com/JuliaInterop/MATLAB.jl}.

\bibitem[{Moore(1998)}]{moore1998cramming}
Moore GE (1998) Cramming more components onto integrated circuits.
\newblock \emph{Proceedings of the IEEE} 86(1): 82--85.

\bibitem[{Moses and Churavy(2020)}]{moses2020instead}
Moses W and Churavy V (2020) Instead of rewriting foreign code for machine
  learning, automatically synthesize fast gradients.
\newblock \emph{Advances in neural information processing systems} 33:
  12472--12485.

\bibitem[{Moses et~al.(2021)Moses, Churavy, Paehler, Hückelheim, Narayanan,
  Schanen and Doerfert}]{Moses2021-ai}
Moses WS, Churavy V, Paehler L, Hückelheim J, Narayanan SHK, Schanen M and
  Doerfert J (2021) Reverse-mode automatic differentiation and optimization of
  {GPU} kernels via enzyme.
\newblock In: \emph{Proceedings of the International Conference for High
  Performance Computing, Networking, Storage and Analysis}. New York, NY, USA:
  ACM.
\newblock \doi{10.1145/3458817.3476165}.
\newblock To be published.

\bibitem[{Munshi(2009)}]{munshi2009opencl}
Munshi A (2009) The {OpenCL} specification.
\newblock In: \emph{2009 IEEE Hot Chips 21 Symposium (HCS)}. IEEE, pp. 1--314.

\bibitem[{Norton et~al.(2022)Norton, Qi and contributors}]{clang_jl}
Norton I, Qi Y and contributors (2022) {Clang.jl}: {J}ulia interface to
  libclang.
\newblock \url{https://github.com/JuliaInterop/Clang.jl}.

\bibitem[{Numrich and Reid(1998)}]{10.1145/289918.289920}
Numrich RW and Reid J (1998) Co-{A}rray {F}ortran for {P}arallel {P}rogramming.
\newblock \emph{SIGPLAN Fortran Forum} 17(2): 1–31.
\newblock \doi{10.1145/289918.289920}.
\newblock \urlprefix\url{https://doi.org/10.1145/289918.289920}.

\bibitem[{{Oak Ridge Leadership Computing Facility}()}]{OLCF}
{Oak Ridge Leadership Computing Facility} (????) {Oak Ridge National
  Laboratory}.
\newblock \urlprefix\url{https://www.olcf.ornl.gov}.

\bibitem[{Omlin and R{\"a}ss(2019)}]{parallelstencil_jl}
Omlin S and R{\"a}ss L (2019) {ParallelStencil.jl}: Package for writing
  high-level code for parallel high-performance stencil computations that can
  be deployed on both {GPU}s and {CPU}s.
\newblock \url{https://github.com/omlins/ParallelStencil.jl}.

\bibitem[{Omlin et~al.(2020)Omlin, R{\"{a}}ss, Kwasniewski, Malvoisin and
  Podladchikov}]{juliacon2020scaling}
Omlin S, R{\"{a}}ss L, Kwasniewski G, Malvoisin B and Podladchikov YY (2020)
  {Solving Nonlinear Multi-Physics on GPU Supercomputers with Julia}.
\newblock \url{https://youtu.be/vPsfZUqI4_0}. Talk presented at JuliaCon.

\bibitem[{Omlin et~al.(2019)Omlin, R{\"a}ss and Utkin}]{implicitglobalgrid_jl}
Omlin S, R{\"a}ss L and Utkin I (2019) {ImplicitGlobalGrid.jl}: Almost trivial
  distributed parallelization of stencil-based {GPU} and {CPU} applications on
  a regular staggered grid.
\newblock \url{https://github.com/eth-cscs/ImplicitGlobalGrid.jl}.

\bibitem[{Orban et~al.(2020)Orban, Siqueira and
  {contributors}}]{orban-siqueira-linearoperators-2020}
Orban D, Siqueira AS and {contributors} (2020) {LinearOperators.jl}.
\newblock https://github.com/JuliaSmoothOptimizers/LinearOperators.jl.
\newblock \doi{10.5281/zenodo.2559295}.

\bibitem[{Ousterhout(1998)}]{Ousterhout1998-ac}
Ousterhout JK (1998) Scripting: higher level programming for the 21st century.
\newblock \emph{Computer} 31(3): 23--30.
\newblock \doi{10.1109/2.660187}.

\bibitem[{Parashar et~al.(1994)Parashar, Hariri, Haupt and
  Fox}]{10.1007/978-3-0348-8534-8_11}
Parashar M, Hariri S, Haupt T and Fox G (1994) A study of software development
  for high performance computing.
\newblock In: Decker KM and Rehmann RM (eds.) \emph{Programming Environments
  for Massively Parallel Distributed Systems}. Basel: Birkh{\"a}user Basel.
\newblock ISBN 978-3-0348-8534-8, pp. 107--116.

\bibitem[{Paszke et~al.(2019)Paszke, Gross, Massa, Lerer, Bradbury, Chanan,
  Killeen, Lin, Gimelshein, Antiga et~al.}]{paszke2019pytorch}
Paszke A, Gross S, Massa F, Lerer A, Bradbury J, Chanan G, Killeen T, Lin Z,
  Gimelshein N, Antiga L et~al. (2019) Pytorch: An imperative style,
  high-performance deep learning library.
\newblock \emph{Advances in neural information processing systems} 32.

\bibitem[{Pich et~al.(2019)Pich, Mui{\~n}os, Lolkema, Steeghs, Gonzalez-Perez
  and Lopez-Bigas}]{pich2019mutational}
Pich O, Mui{\~n}os F, Lolkema MP, Steeghs N, Gonzalez-Perez A and Lopez-Bigas N
  (2019) The mutational footprints of cancer therapies.
\newblock \emph{Nature genetics} 51(12): 1732--1740.
\newblock \doi{10.1038/s41588-019-0525-5}.

\bibitem[{Plietzsch et~al.(2022)Plietzsch, Kogler, Auer, Merino, Gil-de Muro,
  Li{\ss}e, Vogel and Hellmann}]{plietzsch2022powerdynamics}
Plietzsch A, Kogler R, Auer S, Merino J, Gil-de Muro A, Li{\ss}e J, Vogel C and
  Hellmann F (2022) {PowerDynamics.jl} - {A}n experimentally validated
  open-source package for the dynamical analysis of power grids.
\newblock \emph{SoftwareX} 17: 100861.
\newblock \doi{10.1016/j.softx.2021.100861}.

\bibitem[{Rackauckas and Nie(2017)}]{rackauckas2017differentialequations}
Rackauckas C and Nie Q (2017) {DifferentialEquations.jl} {--} {A} performant
  and feature-rich ecosystem for solving differential equations in {J}ulia.
\newblock \emph{Journal of Open Research Software} 5(1): 15.
\newblock \doi{10.5334/jors.151}.

\bibitem[{Rackauckas and Nie(2019)}]{rackauckas2019confederated}
Rackauckas C and Nie Q (2019) Confederated modular differential equation {API}s
  for accelerated algorithm development and benchmarking.
\newblock \emph{Advances in Engineering Software} 132: 1--6.
\newblock \doi{10.1016/j.advengsoft.2019.03.009}.

\bibitem[{Rackauckas et~al.(2020)Rackauckas, Singhvi, Ma, Hatherly, Jones,
  Caine, Saba, TagBot and Olver}]{rackauckas2020sciml}
Rackauckas C, Singhvi A, Ma Y, Hatherly M, Jones S, Caine C, Saba E, TagBot J
  and Olver S (2020) Sciml/differentialequations. jl: v6. 15.0 .

\bibitem[{Ramadhan et~al.(2020)Ramadhan, Wagner, Hill, Campin, Churavy, Besard,
  Souza, Edelman, Ferrari and Marshall}]{ramadhan2020oceananigans}
Ramadhan A, Wagner G, Hill C, Campin JM, Churavy V, Besard T, Souza A, Edelman
  A, Ferrari R and Marshall J (2020) Oceananigans.jl: {F}ast and friendly
  geophysical fluid dynamics on {GPU}s.
\newblock \emph{Journal of Open Source Software} 5(53).
\newblock \doi{10.21105/joss.02018}.

\bibitem[{Ranocha et~al.(2022)Ranocha, Schlottke-Lakemper, Winters, Faulhaber,
  Chan and Gassner}]{ranocha2021adaptive}
Ranocha H, Schlottke-Lakemper M, Winters AR, Faulhaber E, Chan J and Gassner G
  (2022) Adaptive numerical simulations with {T}rixi.jl: {A} case study of
  {J}ulia for scientific computing.
\newblock \emph{Proceedings of the JuliaCon Conferences} 1(1): 77.
\newblock \doi{10.21105/jcon.00077}.

\bibitem[{R{\"{a}}ss et~al.(2019)R{\"{a}}ss, Omlin and Podladchikov}]{gtc2019}
R{\"{a}}ss L, Omlin S and Podladchikov YY (2019) {Resolving Spontaneous
  Nonlinear Multi-Physics Flow Localization in 3-D: Tackling Hardware Limit}.
\newblock \urlprefix\url{https://developer.nvidia.com/gtc/2019/video/S9368}.
\newblock {GTC Silicon Valley - 2019}.

\bibitem[{R\"ass et~al.(2022)R\"ass, Utkin, Duretz, Omlin and
  Podladchikov}]{rass2022assessing}
R\"ass L, Utkin I, Duretz T, Omlin S and Podladchikov YY (2022) Assessing the
  robustness and scalability of the accelerated pseudo-transient method.
\newblock \emph{Geoscientific Model Development} 15(14): 5757--5786.
\newblock \doi{10.5194/gmd-15-5757-2022}.
\newblock \urlprefix\url{https://gmd.copernicus.org/articles/15/5757/2022/}.

\bibitem[{Reed et~al.(2022)Reed, Gannon and
  Dongarra}]{https://doi.org/10.48550/arxiv.2203.02544}
Reed D, Gannon D and Dongarra J (2022) Reinventing {H}igh {P}erformance
  {C}omputing: {C}hallenges and {O}pportunities.
\newblock \doi{10.48550/ARXIV.2203.02544}.
\newblock \urlprefix\url{https://arxiv.org/abs/2203.02544}.

\bibitem[{Regier et~al.(2018)Regier, Pamnany, Fischer, Noack, Lam, Revels,
  Howard, Giordano, Schlegel, McAuliffe et~al.}]{regier2018cataloging}
Regier J, Pamnany K, Fischer K, Noack A, Lam M, Revels J, Howard S, Giordano R,
  Schlegel D, McAuliffe J et~al. (2018) Cataloging the visible universe through
  {B}ayesian inference in {J}ulia at petascale.
\newblock In: \emph{2018 IEEE International Parallel and Distributed Processing
  Symposium (IPDPS)}. IEEE, pp. 44--53.
\newblock \doi{10.1016/j.jpdc.2018.12.008}.

\bibitem[{{Revels} et~al.(2016){Revels}, {Lubin} and
  {Papamarkou}}]{revels2016forward}
{Revels} J, {Lubin} M and {Papamarkou} T (2016) Forward-{M}ode {A}utomatic
  {D}ifferentiation in {J}ulia.

\bibitem[{Reyes and Lom{\"u}ller(2016)}]{reyes2016sycl}
Reyes R and Lom{\"u}ller V (2016) {SYCL}: {S}ingle-source {C++} accelerator
  programming.
\newblock In: \emph{Parallel Computing: On the Road to Exascale}. IOS Press,
  pp. 673--682.

\bibitem[{Rowley(2022)}]{PythonCall.jl}
Rowley C (2022) Pythoncall.jl: Python and julia in harmony.
\newblock \urlprefix\url{https://github.com/cjdoris/PythonCall.jl}.

\bibitem[{Rutkowski(2022)}]{cbinding_jl}
Rutkowski K (2022) {CBinding.jl}: {A}utomatic {C} interfacing for {J}ulia.
\newblock \url{https://github.com/analytech-solutions/CBinding.jl}.

\bibitem[{Saba and contributors(2022)}]{binaybuilder_jl}
Saba E and contributors (2022) {BinaryBuilder.jl}: {B}inary dependency builder
  for {J}ulia.
\newblock \url{https://github.com/JuliaPackaging/BinaryBuilder.jl}.

\bibitem[{Samaroo et~al.(2013)Samaroo, Besard, Churavy, Lin and other
  contributors}]{amdgpu_jl}
Samaroo J, Besard T, Churavy V, Lin D and other contributors (2013)
  {AMDGPU.jl}: {AMD} {GPU} ({ROC}m) programming in {J}ulia.
\newblock \url{https://github.com/JuliaGPU/AMDGPU.jl}.

\bibitem[{Saraswat et~al.(2007)Saraswat, Sarkar and von
  Praun}]{10.1145/1229428.1229483}
Saraswat VA, Sarkar V and von Praun C (2007) X10: {C}oncurrent {P}rogramming
  for {M}odern {A}rchitectures.
\newblock In: \emph{Proceedings of the 12th ACM SIGPLAN Symposium on Principles
  and Practice of Parallel Programming}, PPoPP '07. New York, NY, USA:
  Association for Computing Machinery.
\newblock ISBN 9781595936028, p. 271.
\newblock \doi{10.1145/1229428.1229483}.
\newblock \urlprefix\url{https://doi.org/10.1145/1229428.1229483}.

\bibitem[{Sch\"afer et~al.(2021)Sch\"afer, Tarek, White and
  Rackauckas}]{schafer2021abstractdifferentiation}
Sch\"afer F, Tarek M, White L and Rackauckas C (2021)
  {AbstractDifferentiation.jl}: Backend-agnostic differentiable programming in
  {J}ulia.

\bibitem[{Schlottke-Lakemper et~al.(2021)Schlottke-Lakemper, Winters, Ranocha
  and Gassner}]{schlottkelakemper2021purely}
Schlottke-Lakemper M, Winters AR, Ranocha H and Gassner GJ (2021) A purely
  hyperbolic discontinuous {G}alerkin approach for self-gravitating gas
  dynamics.
\newblock \emph{Journal of Computational Physics} :
  110467\doi{10.1016/j.jcp.2021.110467}.

\bibitem[{Schnetter and contributors(2016)}]{simd_jl}
Schnetter E and contributors (2016) {SIMD.jl}: Explicit {SIMD} vector
  operations for {J}ulia.
\newblock \url{https://github.com/eschnett/SIMD.jl}.

\bibitem[{Scholtz and Wiedenbeck(1990)}]{ScholtzWiedenbeck90}
Scholtz J and Wiedenbeck S (1990) Learning second and subsequent programming
  languages: {A} problem of transfer.
\newblock \emph{International Journal of Human–Computer Interaction} 2(1):
  51--72.
\newblock \doi{10.1080/10447319009525970}.

\bibitem[{Shalf and Leland(2015)}]{7368023}
Shalf JM and Leland R (2015) Computing beyond {M}oore's {L}aw.
\newblock \emph{Computer} 48(12): 14--23.
\newblock \doi{10.1109/MC.2015.374}.

\bibitem[{Shang et~al.(2022)Shang, Shen, Fan, Xu, Guo, Liu, Zhou, Ma, Lin,
  Yang, Li, Wang, Zhang and Li}]{Shang2022-pq}
Shang H, Shen L, Fan Y, Xu Z, Guo C, Liu J, Zhou W, Ma H, Lin R, Yang Y, Li F,
  Wang Z, Zhang Y and Li Z (2022) {Large-Scale} simulation of quantum
  computational chemistry on a new sunway supercomputer .

\bibitem[{Shrestha et~al.(2020)Shrestha, Botta, Barik and
  Parnin}]{ShresthaBottaEtAl20}
Shrestha N, Botta C, Barik T and Parnin C (2020) Here {W}e {G}o {A}gain: {W}hy
  {I}s {I}t {D}ifficult for {D}evelopers to {L}earn {A}nother {P}rogramming
  {L}anguage?
\newblock In: \emph{2020 IEEE/ACM 42nd International Conference on Software
  Engineering (ICSE)}. pp. 691--701.

\bibitem[{Stevens et~al.(2020)Stevens, Taylor, Nichols, Maccabe, Yelick and
  Brown}]{osti_1604756}
Stevens R, Taylor V, Nichols J, Maccabe AB, Yelick K and Brown D (2020) {AI}
  for {S}cience: {R}eport on the {D}epartment of {E}nergy ({DOE}) {T}own
  {H}alls on {A}rtificial {I}ntelligence ({AI}) for {S}cience
  \doi{10.2172/1604756}.
\newblock \urlprefix\url{https://www.osti.gov/biblio/1604756}.

\bibitem[{Straßel et~al.(2020)Straßel, Reusch and Keuper}]{9307940}
Straßel D, Reusch P and Keuper J (2020) Python workflows on hpc systems.
\newblock In: \emph{2020 IEEE/ACM 9th Workshop on Python for High-Performance
  and Scientific Computing (PyHPC)}. pp. 32--40.
\newblock \doi{10.1109/PyHPC51966.2020.00009}.

\bibitem[{Stroustrup(2013)}]{stroustrup2013c++}
Stroustrup B (2013) \emph{The {C++} programming language}.
\newblock Pearson Education.

\bibitem[{{The HDF Group}(2000-2010)}]{hdf5}
{The HDF Group} (2000-2010) {Hierarchical data format version 5}.
\newblock \urlprefix\url{http://www.hdfgroup.org/HDF5}.

\bibitem[{van~der Plas et~al.(2022)van~der Plas, Dral, Berg,
  {\textPi\textalpha\textnu\textalpha\textgamma\textiota\textomega\texttau\texteta\textvarsigma}~{\textGamma\textepsilon\textomega\textrho\textgamma\textalpha\textkappa\textomikron\textpi\textomikron\textupsilon\textlambda\textomikron\textvarsigma},
  Bochenski, disberd, Lungwitz, Huijzer, Zhang, Schneider, Weaver, Rogerluo,
  Kadowaki, Ling, Wu, Burns, Gerritsen, Novosel, Supanat, Moon, pupuis, Abbott,
  Bauer, Bouffard, Terasaki, Polasa, TheCedarPrince and fghzxm}]{pluto}
van~der Plas F, Dral M, Berg P,
  {\textPi\textalpha\textnu\textalpha\textgamma\textiota\textomega\texttau\texteta\textvarsigma}~{\textGamma\textepsilon\textomega\textrho\textgamma\textalpha\textkappa\textomikron\textpi\textomikron\textupsilon\textlambda\textomikron\textvarsigma},
  Bochenski N, disberd, Lungwitz B, Huijzer R, Zhang E, Schneider FSS, Weaver
  I, Rogerluo, Kadowaki S, Ling J, Wu Z, Burns C, Gerritsen J, Novosel R,
  Supanat, Moon Z, pupuis, Abbott M, Bauer N, Bouffard P, Terasaki S, Polasa S,
  TheCedarPrince and fghzxm (2022) fonsp/pluto.jl: v0.17.7.
\newblock \doi{10.5281/zenodo.5889169}.
\newblock \urlprefix\url{https://doi.org/10.5281/zenodo.5889169}.

\bibitem[{Verdugo and Badia(2021)}]{verdugo2021software}
Verdugo F and Badia S (2021) The software design of {G}ridap: a finite element
  package based on the {J}ulia {JIT} compiler.

\bibitem[{Vetter et~al.(2018)Vetter, Brightwell, Gokhale, McCormick, Ross,
  Shalf, Antypas, Donofrio, Humble, Schuman, Van~Essen, Yoo, Aiken, Bernholdt,
  Byna, Cameron, Cappello, Chapman, Chien, Hall, Hartman-Baker, Lan, Lang,
  Leidel, Li, Lucas, Mellor-Crummey, Peltz~Jr., Peterka, Strout and
  Wilke}]{osti_1473756}
Vetter JS, Brightwell R, Gokhale M, McCormick P, Ross R, Shalf J, Antypas K,
  Donofrio D, Humble T, Schuman C, Van~Essen B, Yoo S, Aiken A, Bernholdt D,
  Byna S, Cameron K, Cappello F, Chapman B, Chien A, Hall M, Hartman-Baker R,
  Lan Z, Lang M, Leidel J, Li S, Lucas R, Mellor-Crummey J, Peltz~Jr P, Peterka
  T, Strout M and Wilke J (2018) {Extreme {H}eterogeneity 2018 - {P}roductive
  {C}omputational {S}cience in the {E}ra of {E}xtreme {H}eterogeneity: {R}eport
  for {DOE} {ASCR} {W}orkshop on {E}xtreme {H}eterogeneity}
  \doi{10.2172/1473756}.

\bibitem[{Weitz et~al.(2020)Weitz, Beckett, Coenen, Demory, Dominguez-Mirazo,
  Dushoff, Leung, Li, M{\u{a}}g{\u{a}}lie, Park et~al.}]{weitz2020modeling}
Weitz JS, Beckett SJ, Coenen AR, Demory D, Dominguez-Mirazo M, Dushoff J, Leung
  CY, Li G, M{\u{a}}g{\u{a}}lie A, Park SW et~al. (2020) Modeling shield
  immunity to reduce {COVID}-19 epidemic spread.
\newblock \emph{Nature medicine} 26(6): 849--854.
\newblock \doi{10.1038/s41591-020-0895-3}.

\bibitem[{Wienke et~al.(2012)Wienke, Springer, Terboven
  et~al.}]{wienke2012openacc}
Wienke S, Springer P, Terboven C et~al. (2012) Open{ACC}—first experiences
  with real-world applications.
\newblock In: \emph{European Conference on Parallel Processing}. Springer, pp.
  859--870.

\bibitem[{Williams(2016)}]{theregister:left-pad}
Williams C (2016) How one developer just broke {N}ode, {B}abel and thousands of
  projects in 11 lines of {J}ava{S}cript.
\newblock
  \urlprefix\url{https://www.theregister.com/2016/03/23/npm_left_pad_chaos/}.

\bibitem[{Zhu et~al.(2021)Zhu, AlAwar, Erez and Gligoric}]{9678726}
Zhu S, AlAwar N, Erez M and Gligoric M (2021) Dynamic generation of python
  bindings for hpc kernels.
\newblock In: \emph{2021 36th IEEE/ACM International Conference on Automated
  Software Engineering (ASE)}. pp. 92--103.
\newblock \doi{10.1109/ASE51524.2021.9678726}.

\end{thebibliography}

\ifdraft{
\newpage
\tableofcontents
\newpage
\listoftodos[Summary of Embedded Notes]
}

\end{document}